\documentclass[nofootinbib,floatfix,preprint]{revtex4}

\usepackage{epsfig}
\usepackage{graphicx}

\newcommand{\be}{\begin{equation}}
\newcommand{\ee}{\end{equation}}

\begin{document}

\title{A model of diffusion in a potential well for the dynamics of the large-scale circulation in turbulent Rayleigh-B{\'e}nard convection}
\author{Eric Brown}
\altaffiliation[Present address:  ]{The James Franck Institute and Department of Physics, The University of Chicago, Chicago, IL 60637}
\author{Guenter Ahlers}
\affiliation{Department of Physics and iQCD, University of California, Santa Barbara, CA 93106}
\date{\today}
 
\begin{abstract}
Experimental measurements of properties of the large-scale circulation (LSC) in turbulent convection of a fluid heated from below in a cylindrical container of aspect ratio one are presented and used to test a model of diffusion in a potential well for the  LSC.  The model  consists of a pair of stochastic ordinary differential equations motivated by the Navier-Stokes equations. The two coupled equations are for the azimuthal orientation $\theta_0$, and for the azimuthal temperature amplitude $\delta$ at the horizontal midplane. The dynamics is due to the driving by Gaussian distributed white noise that is introduced to represent the action of the small-scale turbulent fluctuations on the large-scale flow. Measurements of the diffusivities that determine the noise intensities are reported. Two time scales predicted by the model are found to be within a factor of two or so of corresponding experimental measurements. A scaling relationship predicted by the model between $\delta$ and the Reynolds number is confirmed by measurements over a large experimental parameter range.  The Gaussian peaks of probability distributions $p(\delta)$ and $p(\dot\theta_0)$ are accurately described by the model; however the non-Gaussian tails of $p(\delta)$ are not.    The frequency, angular change, and amplitude bahavior during cessations are accurately described by the model when the tails of the probability distribution of $\delta$ are used as experimental input. 
\end{abstract}

 \maketitle

\section{Introduction}

Turbulent convection is one of the outstanding unsolved problems of classical physics (for reviews, see for example Refs. \cite{Si94, Ka01, AGL02}).  The problem of Rayleigh-B{\'e}nard convection (RBC) consists of a fluid sample heated from below.  The heat input causes the emission of volumes of hot fluid known as ``plumes" from  a bottom thermal boundary layer that rise due to buoyancy, while cold plumes emitted from a top boundary layer sink.  The experiments are done in cylindrical containers with an aspect ratio $\Gamma \equiv D/L \approx 1$ ($L$ is the height and $D$ is the diameter of the sample).  In the turbulent regime of $\Gamma=1$ samples, the plumes drive a large-scale circulation (LSC), also known as the ``mean wind",  which is oriented nearly vertically with up-flow and down-flow on opposite sides of the sample \cite{KH81, SWL89, CGHKLTWZZ89, CCL96, QT01a, FA04, SXT05,TMMS05}. The wind in turn carries the plumes, primarily  up one side and down the other. The dynamics of the LSC include oscillations \cite{HCL87,CGHKLTWZZ89,CCL96,TSGS96,CCS97,QYT00,QT01a,QT01b, NSSD01, QT02, QSTX04} in which the orientation of the upper half of the LSC oscillates out of phase with the lower half \cite{FA04,FBA08}.  The LSC breaks the rotational symmetry of a cylindrical sample and its circulation plane must somehow choose an azimuthal orientation.  This orientation has been found to undergo spontaneous diffusive meandering \cite{SXX05, XZX06, BA06a, BA06b}.   The LSC also undergoes re-orientations both by azimuthal {\em rotations} \cite{CCS97, BA06a}, and by {\em cessations} in which the LSC slows to a stop and restarts in a random new orientation \cite{BNA05,BA06a}.   On longer time scales, Earth's Coriolis force was found to cause a net rotation of the LSC orientation on average once every 3 days, and to align the LSC in a preferred orientation close to West \cite{BA06b}.  These LSC dynamics observed in experiments may be related to some natural convection dynamics.  For example, it is possible that cessations of the flow in the Earth's outer core are responsible for changes in the orientation of Earth's magnetic field \cite{GCHR99}.   Reversals are known to occur in the wind direction in the atmosphere \cite{DDSC00}.  Torsional oscillations are observed in the solar convection zone \cite{HL80b}.

Two stochastic models of LSC flow-reversal have been proposed in the literature \cite{SBN02, Be05}.  They treated  diffusion of the LSC strength in a potential well, but there was no physical motivation for the shape of the potential that was used  (which differs from ours) and the model parameters were chosen phenomenologically. No azimuthal degree of freedom was included, and thus only genuine reversals  of the LSC (which are now known to be very rare events) could be produced. Two other models \cite{FGL05, RPTDGFL06} describe the LSC with deterministic differential equations that have  chaotic  solutions. They have their roots in the Navier-Stokes equations and retain terms that are argued to be physically important. One of them \cite{FGL05} again is lacking the azimuthal degree of freedom that is so important to the LSC dynamics found in experiment. The other \cite{RPTDGFL06} is based on an exact solution of the Boussinesq equations in the inviscid and unforced limit, but employs physically unrealistic boundary conditions and adds dissipation {\it a posteriori}. It lacks the stochastic character found in experiment, and requires the arbitrary adjustment of parameters to yield the chaotic solutions that might be compared with observations.

In order to improve upon the state of the field described above, we presented briefly  in Ref.~\cite{BA07a} a stochastic model of the LSC that was motivated by the physically relevant terms of the Navier-Stokes (NS)  equations for RBC. It is in the same spirit as a model for the effects of Earth's Coriolis force on the flow \cite{BA06b}, and we will show in a subsequent paper that the Coriolis-force model is consistent with the strong-damping limit of the current model.  The model consists of two coupled stochastic ordinary differential equations (ODEs): one for the strength of the LSC represented by an amplitude $\delta$ of the azimuthal temperature variation at the horizontal mid-plane of the sample, and the other for the azimuthal LSC orientation $\theta_0$. For $\delta$ it leads to diffusive motion in a potential which has a minimum at $\delta_0 > 0$ and a maximum at $\delta = 0$.  On the rare occasions when the diffusion reaches (or comes close to)  the maximum, then a cessation has occurred.  The shape of that potential follows from taking a volume average of contributions from buoyancy and from the viscous drag on the walls. The equation for $\theta_0$ contains a nonlinear coupling to the $\delta$-equation which is proportional to $\delta$, which comes from the advective term in the NS equation, and which represents the angular momentum of the circulation. Thus, when $\delta$ is large, the angular momentum is large and the LSC orientation is relatively immune to re-orientation by the stochastic forces. On the other hand, near cessations where $\delta$ becomes small, re-orientations are relatively easily achieved by the fluctuating background. 
The stochastic driving represents phenomenologically the action of the small-scale turbulent fluctuations on the large-scale circulation. The strength of these fluctuating forces is obtained from experiment by measuring the diffusivities of the relevant variables.
The model was solved numerically, and the numerical solutions were found to reproduce quite well the cessations and rotations, as well as the diffusive azimuthal meandering,  that had been observed in experiments \cite{CCS97,BNA05,BA06a}. The model also yielded probability distributions for the angular displacement during cessations and rotations that were in quite good agreement with the measurements \cite{BNA05,BA06a}. 

In the present paper we present a more detailed derivation of and motivation for the model. 
Then we derive new analytic results by a lowest-order expansion of the potential for the $\delta$ equation about its minimum, and by linearizing the coupling term of the $\theta_0$ equation. We present experimental measurements of the model parameters, as well as of the  Rayleigh-number dependences of several quantities, and for the most part find reasonable agreement of these results with the model predictions. A notable exception is the shape of the $\delta$-potential near $\delta = 0$. We suggest that a possible explanation for this disagreement may be found in the neglect by the model of diffusive heat transport across the top and bottom boundary layers which can be expected to become significant when the LSC amplitude becomes small. 

In the next section we discuss the experiment. This is kept brief because much of it was done before \cite{BNFA05, BA06a}. In Sect.~\ref{sec:derivation} we give a detailed derivation of and motivation for the model. A linearized version of the model is derived in Sect.~\ref{sec:linear}. The potential for the $\delta$ equation, and its linearized version, are discussed in Sect.~\ref{sec:potential}. In Sect.~\ref{sect:parameters} we present new experimental measurements of the model parameters. First  measurements of the Reynolds numbers are discussed. Then the  diffusivities corresponding to stochastic fluctuations are presented. Next, probability distributions, power spectra, and correlations of $\delta$ and  $\dot \theta_0$ are given. This is followed in Sect~\ref{sect:scaling} by a comparison between model predictions for various Rayleigh-number dependences and measurements of the model time scales, of $\delta_0$, of the diffusivities, and of further interesting dimensionless parameters. 
In Sect.~\ref{sect:cessations} we use the experimental measurement of the tails of the probability distribution of $\delta$ as an experimental input to the model to predict the behavior of cessations more accurately.  We predict the duration, frequency, and net angular change during cessations and compare these values to the experimental results of Ref.~\cite{BA06a}. In Sect.~\ref{sec:expanded_model} we suggest a modification of the model that would add a third equation and that would be expected to account for the small-$\delta$ behavior without the empirical experimental input.

\section{The experiment}

In order to measure the model parameters, experimental data from two cylindrical samples with aspect ratio $\Gamma \approx 1$ were used. These were the medium and large sample described in detail elsewhere \cite{BNFA05, BA06a}.  The samples had copper top and bottom plates and a plexiglas side wall.  The medium sample had a diameter $D=24.81$ cm and a height $L=24.76$ cm, and the large sample had $D = 49.67$ cm and $L = 50.61$ cm.  The Rayleigh number $R$ is given by

\be
R \equiv \frac{\alpha g \Delta T L^3}{\kappa \nu}
\label{eq:rayleigh}
\ee

\noindent where $\alpha$ is the isobaric thermal expansion coefficient, $g$ the acceleration of gravity, $\Delta T$ the applied temperature difference, $\kappa$ the thermal diffusivity, and $\nu$ the kinematic viscosity.  By applying a temperature difference $0.5 \mbox{ K } < \Delta T < 20$ K between the bottom and top plates, with two samples of different heights $L$,  a Rayleigh number range of $3\times10^8 \stackrel{<}{_\sim} R \stackrel{<}{_\sim} 10^{11}$ could be covered.  The Prandtl number $\sigma$ is given by

\be
\sigma \equiv \frac{\nu}{\kappa} \ .
\label{eq:prandtl}
\ee

\noindent Each sample was filled with water and the average temperature between the bottom and top plates usually was kept at $40.0^{\circ}$ C, giving  $\sigma = 4.38$.  Some measurements were made at other temperatures and permitted a change of $\sigma$ over the range $3.3 \alt \sigma \alt 5.5$. Measurements were made with thermistors placed into blind holes drilled into the side wall from the outside so they did not interfere with the flow.  There were eight thermistors at the mid-height of the side wall, equally spaced azimuthally.  The LSC carried warm fluid from the bottom plate up one side of the sample, which cooled when it passed the top plate and went down on the opposite side of the sample.  The temperature profile

\be
T = T_0 + \delta\cos(\theta_0-\theta)
\label{eq:temp_profile}
\ee

\noindent was fit to the temperature measurements where $\delta$ characterizes the strength of the LSC and $\theta_0$ is the orientation of the LSC \cite{BA06a,BA07b}.   Fits of the temperature measurements every 2.5 seconds provide time series of $\delta$ and $\theta_0$.  These time series contained the diffusive dynamics of the LSC and the re-orientations \cite{BNA05, BA06a}.  The model parameters can be extracted from the time series as described in Sect.~\ref{sect:parameters}.

\section{The model}
\label{sec:model}

\subsection{Derivation}
\label{sec:derivation}
	
The model presented in Ref.~\cite{BA07a} is reproduced here in more detail.  The LSC strength can be described by the velocity component $u_{\phi}$. Here $\phi$ is an angle in the vertical circulation plane of the LSC as shown in Fig.~\ref{fig:wind_phi}, and $u_{\phi}$ describes the flow in the absence of azimuthal motion.  One expects the acceleration to be due to a balance between buoyancy and drag forces.  The pressure term primarily provides the inward force to keep the LSC in a loop, and it is not expected to contribute a net force in the $\phi$-direction.  Thus only the buoyancy and viscous drag terms are included on the right hand side of the NS equation for $u_{\phi}$, and we neglect the nonlinear term:

\be
\dot u_{\phi}  = g\alpha (T-T_0) + \nu\nabla^2 u_{\phi} \ .
\label{eq:nav}
\ee

\noindent  To obtain a model in the form of an ODE that describes the flow with only a few variables, a global average is taken over the field variables that retains the essential physics of the LSC. This average can be carried out using the experimental observation \cite{BNA05, ABN06, BA06a, BA06b} that the temperature of the LSC at the side wall at mid-height is given by Eq.~\ref{eq:temp_profile}.   The profile is interpolated to be 

\be
T(r,\theta) = T_0 + \frac{2r\delta}{L}\cos(\theta_0 - \theta) 
\label{eq:temp_profile_r}
\ee

\noindent where $r$ is the radius measured from the cylinder axis.  An approximately linear radial variation was found experimentally in Ref.   \cite{QT01a}.  The buoyancy acts on the entire LSC. It enhances the LSC in proportion to its vertical component. Thus, to approximately account for this, we multiply  the volume average by  a factor of $1/2$. The buoyancy term can now be approximated by an integral over the container volume $V$ using Eq.~\ref{eq:temp_profile_r} to obtain $1/(2V)\int_V g\alpha(T-T_0)S(\theta-\theta_0)dV = 2g\alpha\delta/(3\pi)$ where $S$ is a step function with $S = 1$ for $|\theta_0 - \theta| < \pi/2$ and $S = -1$ otherwise.   Note that when the volume averages are taken, assumptions about the geometry of the flow only affect coefficients by factors of order one, and do not influence the functional form of the results.

It was argued in Ref.~\cite{BA06a} based on the experiments of Ref.~\cite{SXT05} that the azimuthal velocity profile is  close to a step function, so the bulk velocity profile is assumed to be 
\be
u_{\phi}(r,\theta) = \frac{2rU}{L}  
\label{eq:velocity_profile}
\ee

\noindent where $U$ is the maximum vertical speed near the side wall. Here again the linear variation with $r$ is supported by experimental results \cite{QT01a,QT01b}. The drag is assumed to occur in the viscous boundary layers, where $\nabla^2 u_{\phi} \approx -U/\lambda^2$ ($\lambda$ is the viscous  boundary-layer width).  The viscous boundary layers on the side wall and plates occupy a fraction of the container equal to $6\lambda/L$, so the volume average of the drag is $-6\nu U/(\lambda L)$.  The viscous boundary-layer width is assumed to follow the Prandtl-Blasius form $\lambda = L R_{e,i}^{-1/2}/2$ with a fluctuating Reynolds number $R_{e,i} \equiv  U L/\nu$.  Although this must be regarded as an approximation, the Prandtl-Blasius form for the boundary layer has worked remarkably well in previous models (for example, \cite{BA06b}). It also has been very successful in predicting the dependence of the Reynolds number on the Rayleigh number \cite{GL02}, and in treating non-Boussinesq effects on the Nusselt number and the center temperature \cite{ABFFGL06,AFFGL07}. With this form the damping term becomes nonlinear in $U$.  A volume average of the acceleration term using Eq.~\ref{eq:velocity_profile} results in  $(1/V)\int_V \dot u_{\phi}dV = 2\dot U/3$.  Combining these results gives the volume-averaged equation

\be
\frac{2\dot U}{3} = \frac{2g\alpha\delta}{3\pi} - \frac{12 \nu^{1/2} U^{3/2}}{L^{3/2}} \ .
\label{eq:uave}
\ee

\noindent This equation has two variables, $\delta$ and $U$, but we only have experimental measurements of $\delta$.  In order to compare the model to current data,  it is assumed that the temperature amplitude $\delta$ is instantaneously proportional to the speed $U$, since both variables are measures of the LSC strength. This assumption is consistent with simultaneous velocity and temperature measurements at the same point at the mid-height near the side wall which gave a correlation of 0.8 \cite{Tong_private}.  Two-dimensional direct numerical simulations also found that on average $\delta$ is proportional to $U$, although instantaneously both fluctuate, with the same correlation of 0.8 \cite{Kaz}.  The proportionality constant relating $\delta$ to $U$ must satisfy the time-average 

\be
\frac{2g\alpha\langle\delta\rangle}{3\pi}  = \frac{12\nu^2 R_e^{3/2}}{L^3}  
\label{eq:delta_u}
\ee

\noindent  of Eq.~\ref{eq:uave}. This fixes the proportionality at 

\be
\frac{2g\alpha\delta}{3\pi}  = \frac{12\nu U R_e^{1/2}}{L^2} \ .
\label{eq:delta_u_prop}
\ee

\noindent Note that Eq.~\ref{eq:delta_u} forces the sum of the powers of $U$ and $R_e$ to be $3/2$, and so the assumption fixes $\delta \propto U R_e^{1/2}$.  Equation \ref{eq:delta_u_prop} is substituted into Eq.~\ref{eq:uave} and a stochastic term $f_{\delta}(t)$ is added to represent the influence of the small-scale turbulent fluctuations on the large-scale flow to get the Langevin equation

\be
\dot\delta = \frac{\delta}{\tau_{\delta}} - \frac{\delta^{3/2}}{ \tau_{\delta}\sqrt{\delta_0}} + f_{\delta}(t)\ .
\label{eq:lang_delta}
\ee

\noindent Using Eqs.~\ref{eq:rayleigh} and \ref{eq:prandtl} one finds the constant

\be
\delta_0 \equiv \frac{18\pi \Delta T \sigma R_e^{3/2}}{R}
\label{eq:delta0}
\ee

\noindent and the time scale

\be
\tau_{\delta} \equiv \frac{L^2}{18\nu R_e^{1/2}} \ .
\label{eq:tau_delta}
\ee

\noindent  The stochastic term is assumed to be Gaussian distributed white noise with zero mean.  These properties will be explored in detail in Sect.~\ref{sect:parameters}.

Equation \ref{eq:lang_delta} has two fixed points, one unstable at $\delta=0$ and one stable when $\delta = \delta_0$. Thus, in the absence of fluctuations, $\delta_0$ can be interpreted as the steady-state amplitude.  The stochastic equation reproduces some of the important behavior of the LSC.  When a temperature difference is applied to generate buoyancy, the LSC will start to grow due to the instability at $\delta = 0$ until it reaches the stable fixed point at $\delta = \delta_0$.  If the fluctuations are small, $\delta$ spends most of its time meandering near the stable fixed point at $\delta_0$, and if the fluctuations are large enough the LSC occasionally undergoes a cessation when fluctuations drive $\delta$ close to $0$.

Other models of the LSC dynamics \cite{SBN02, Be05} have used an equation for the LSC strength similar to Eq.~\ref{eq:lang_delta}, but assumed an exponent of 3 for the damping term instead of $3/2$.  The essential physics of Eq.~\ref{eq:lang_delta} is in the (in)stability behavior of the fixed points.  Equations like Eq.~\ref{eq:lang_delta} have one unstable and one stable fixed point as long as the damping exponent is greater than one. Thus in a qualitative sense Eq.~\ref{eq:lang_delta} has a behavior similar to that of the earlier models.  Since $\delta$ is chosen to be non-negative and reversals are accounted for by a change in orientation\footnote{The parameters of the temperature profile $T= T_0 + \delta\cos(\theta - \theta_0)$ are not uniquely determined, because the change $\delta \rightarrow -\delta$ is equivalent to $\theta_0 \rightarrow \theta_0 \pm \pi$. Thus $\delta$ is chosen to be always non-negative for uniqueness.}, there is no need to restrict the exponents to odd integers as in the other models.  Since the $3/2$ power came from our choice of drag law and scaling relationship between $\delta$ and $U$, the phenomenology of the model is also robust to these choices as long and the exponent of the drag term is greater than that of the buoyancy term.

The second Langevin equation describes the azimuthal motion of the LSC.   The only driving force for azimuthal motion in a symmetric system is turbulent fluctuations, and damping can come from either viscosity or rotational inertia.  Thus only the drag term is kept on the right-hand-side of the NS equation in the azimuthal coordinate:

\be
\dot u_{\theta} + \vec u \cdot \vec\nabla u_{\theta} = \nu\nabla^2 u_{\theta} \ .
\label{eq:nav_theta}
\ee

\noindent  Again, the components of the nonlinear term are negligible except the one corresponding to the rotational inertia of the LSC in the $\phi$-coordinate $(\vec u \cdot \vec\nabla)u_{\theta} \approx (u_{\phi}/r)\partial u_{\theta}/\partial\phi \sim U\dot\theta_0$.  This can be physically understood in terms of the dynamics of a rigid rotator:  a rotating body has some stability due to its angular momentum that resists a torque in an orthogonal direction.    The torque in the $\theta$-direction is equal to $I \ddot\theta_0 = \dot L_{\theta}$ where $L_{\theta}$ and $L_{\phi}$ are the angular momenta in the respective coordinates, and $I$ is the moment of inertia.  Since the LSC nearly fills the container, $I$ is assumed to be the same around both axes of rotation.  For a differential torque applied to a rigid rotator, a change in orientation is more difficult when there is rotation in a perpendicular direction $d\theta_0 = dL_{\theta}/L_{\phi}$, or $\dot L_{\theta} = L_{\phi}\dot \theta_0$, where $L_{\phi} = I u_{\phi}/r \approx 2UI/L$.   Combining these equations yields the inertial contribution to acceleration $\ddot\theta_0 = 2U\dot\theta_0/L$.  Using the approximation that viscous drag occurs mainly in the boundary layers, one has $\langle \nu\nabla^2 u_{\theta} \rangle_V \approx \nu (L\dot\theta_0/2)/\lambda^2 \times 6\lambda/L$.   The volume average of Eq.~\ref{eq:nav_theta} is

\be
\frac{L\ddot\theta_0}{3} = - \frac{2U\dot\theta_0}{3} - \frac{6\nu\dot\theta_0 R_{e,i}^{1/2}}{L}
\label{eq:theta_ave}
\ee

\noindent The ratio of the viscous drag term to the angular-momentum damping is  equal to $9R_{e,i}^{-1/2}$.  At $R_{e,i} = 3700$ (the Reynolds number at $R=1.1\times10^{10}$)  for example, this yields $\approx 0.15$ for this ratio.  Since  rotational inertia  damps the azimuthal motion much more than the viscous drag across the boundary layer near the side wall,  the viscous damping term of the azimuthal equation is neglected from now on.   The azimuthal speed is generally small compared to the LSC speed so the effect of rotational inertia is much larger on the azimuthal coordinate than on the LSC strength, which is why the nonlinear term could be ignored in Eq.~\ref{eq:nav}.    Converting the remaining terms from $U$ to $\delta$ using Eq.~\ref{eq:delta_u_prop}, and adding another stochastic term $f_{\dot\theta}(t)$ representing turbulent fluctuations, gives

\be
\ddot\theta_0 = - \frac{\dot\theta_0\delta}{\tau_{\dot\theta}\delta_0} + f_{\dot\theta}(t)
\label{eq:lang_theta}
\ee

\noindent with the constant time scale

\be
\tau_{\dot\theta} \equiv \frac{L^2}{2\nu R_e} \ .
\label{eq:tau_theta}
\ee

\noindent The two stochastic ODEs Eqs.~\ref{eq:lang_delta} and \ref{eq:lang_theta} compose the model for the LSC dynamics.  The stochastic terms $f_{\delta}(t)$ and $f_{\dot\theta}(t)$ that drive the dynamics of the system are presumed to originate from the small-scale turbulent background fluctuations. We made some extremely simplifying assumptions about them because they cannot be isolated from the dynamical system to independently measure their properties.  They are assumed to be Gaussian distributed, uncorrelated white noise. Then the diffusivities $D_{\delta}$ and $D_{\dot\theta}$ are the only parameters required to describe them, and these can be estimated from experimental data.  The adequacy of these assumptions is tested in the context of the model in Sect.~\ref{sect:parameters}.  There are direct model predictions for $\delta_0$ and the time scales $\tau_{\delta}$ and $\tau_{\dot\theta}$.  Methods for experimentally obtaining all of the parameter values and testing the model will be covered in Sect.~\ref{sect:parameters}.

\subsection{Linear approximation}
\label{sec:linear}

When fluctuations of $\delta$ are small, or $\delta \approx \delta_0$, a linear approximation to the Langevin equations can be made.  For the data of Ref.~\cite{BA06a} this approximation is satisfied most of the time, but not during the quite rare cessations.  Thus the linearized version of the model should apply to long-term averages of data, but the full non-linear model must be used to study cessations.  

Starting with Eq.~\ref{eq:lang_delta}  for $\delta$, we expanded around the stable fixed-point solution $\delta=\delta_0$ by rewriting $\delta = \delta_0 + \epsilon$.   Assuming $\epsilon \ll \delta_0$ and thus keeping up to the first order term in the expansion one has $(\delta_0 + \epsilon)^{3/2} \approx \delta_0^{3/2} + 3\epsilon\sqrt{\delta_0}/2$.  Using $\dot\delta(\delta_0) = 0$, Eq.~\ref{eq:lang_delta} simplifies to a linear equation

\be
\dot\epsilon = -\frac{\epsilon}{2\tau_{\delta}}+ f_{\delta}(t) \ .
\label{eq:lang_delta_lin}
\ee

\noindent Equation \ref{eq:lang_theta} can also be linearized near the stable fixed point by setting $\delta = \delta_0$ to obtain

\be
\ddot\theta_0 = -\frac{\dot\theta_0}{\tau_{\dot\theta}}+ f_{\dot\theta}(t) \ .
\label{eq:lang_theta_lin}
\ee

\noindent  The typical size of fluctuations in $\delta$ can be measured by the ratio of the variance of $\delta$ to the average of $\delta$ squared, or $\sigma_{\delta}^2/\delta_0^2 \approx 0.07$ for the large sample data and smaller in the medium sample (see Sect.~\ref{sect:prob}).  Since this ratio is small, the effect of the variable nature of the damping is on average small.

\subsection{The potential wells}
\label{sec:potential}

It is useful to think of Eq.~\ref{eq:lang_delta} in terms of diffusion in a potential well, as in the Arrhenius-Kramers problem \cite{Kr40}.  The potential is defined by $V_\delta \equiv -\int \dot\delta_d d\delta$ where $\dot\delta_d$ is the deterministic part of Eq.~\ref{eq:lang_delta}. Thus

\be
V_{\delta} =  - \frac{\delta^2}{2\tau_{\delta}} + \frac{2\delta^{5/2}}{5\tau_{\delta}\sqrt{\delta_0}} \ .
\label{eq:potential}
\ee

\noindent This potential is shown in Fig.~\ref{fig:potential}.  Its minimum  is at $\delta = \delta_0$, and due to the stochastic term $f_{\delta}(t)$ the value of $\delta$ fluctuates around $\delta_0$ in the bottom of the well.  A cessation occurs when the LSC amplitude drops to zero, or when fluctutations in $\delta$ cross the potential barrier $\Delta V_{\delta}  \equiv V_{\delta}(0) - V_{\delta}(\delta_0)$.  The linearization of the potential well near the stable fixed point gives

\be
V_{\epsilon} \equiv V_{\delta}(\delta_0) -\int \dot\epsilon_d d\epsilon =  V_{\delta}(\delta_0) +\frac{\epsilon^2}{4\tau_{\delta}} \ ,
\label{eq:potential_lin}
\ee

\noindent where $\dot\epsilon_d$ is the deterministic part of Eq.~\ref{eq:lang_delta_lin}.  This potential is also shown in Fig.~\ref{fig:potential}.  The minima of both potentials overlap and have the same curvature, so the dynamics for small fluctuations will be the same for both; but the non-linear potential $V_{\delta}$ is skewed to bias fluctuations towards small $\delta$ relative to the parabolic potential $V_{\epsilon}$.

Similarly, the dynamics of $\dot\theta_0$, given by Eq.~\ref{eq:lang_theta}, can be thought of as diffusion in the potential well
\be
V_{\dot\theta} \equiv -\int\ddot\theta_0 d\dot\theta_0 = \dot\theta_0^2\delta /(2\delta_0\tau_{\dot\theta}) \ .
\label{eq:potential_theta}
\ee

\noindent In this case the well is parabolic, with a minimum at $\dot \theta = 0$. Thus the mean of $\dot \theta$ will be zero, with fluctuations symmetrically about this value. However, the well curvature varies with $\delta$. Thus, reorientations occur when fluctuations take $\dot\theta_0$ far away from the average of zero, and this is more likely when $\delta$, and thus the curvature of the potential, are small. The extreme case is a cessation, which occurs when $\delta$ is close to zero and the potential $V_{\dot\theta}$ is nearly flat.

\section{Measuring the model parameters}
\label{sect:parameters}
  
\subsection{The Reynolds number $R_e$ and the mean temperature amplitude $\delta_0$}
  
The model of the large-scale circulation requires several input parameters.  The
 Reynolds number $R_e$ was measured in many experiments (for a recent summary,
  see Ref.~\cite{BFA07}) and desribed by theoretical models \cite{GL02}.  Values are
 taken from temperature-correlation functions corresponding to the plume turnover that came from the same apparatus as the current measurements and were reported in Ref.~\cite{BFA07}, and will not be distinguished from different measures of the
 Reynolds number that were discussed in Ref.~\cite{BFA07} because these differences are small compared to the needs of an order-of-magnitude model.  The fixed-point temperature-amplitude $\delta_0$ can be approximated by an average over a time series: $\delta_0 \simeq \langle\delta\rangle$.  Due to the asymmetry of
  Eq.~\ref{eq:potential} around $\delta = \delta_0$, $\langle\delta\rangle$ is slightly less than $\delta_0$, but the difference is a small fraction of $\delta_0$ and so will be ignored.

\subsection{The diffusivities and time scales}

For diffusive fluctuations, the mean-square change $\langle(dx)^2\rangle$ of a variable $x$ over a time interval $dt$ is a linear function of $dt$, and the diffusivity $D_x$ is defined by the equation $\langle(dx)^2\rangle = D_x dt$.  For stepwise numerical simulations, the stochastic terms $f_x(t)$ in the model have a variance $D_x/h$ where $h$ is the time step of the numerical integration.  

\subsubsection{Diffusion of the amplitude $\delta$}

A plot of $\langle(d\delta)^2\rangle \equiv \langle [\delta(t+dt) - \delta(t)]^2\rangle$ as a function of the time interval $dt$ is shown in Fig.~\ref{fig:diff_delta} for $R=1.1\times 10^{10}$ as an example (data at numerous other values of $R$ behave in a similar manner).  The equation $\langle(d\delta)^2\rangle = D_{\delta} dt$ was fit to the data for $30 \mbox{s} < dt < 80$ s to obtain $D_{\delta} = 6.4\cdot 10^{-5} K^2/s$.  Although this linear relationship is characteristic of diffusion, the variance of the change in $\delta$ saturates at a constant value for large $dt$ .  This happens because the diffusion of $\delta$ occurs in the potential well given by Eq.~\ref{eq:potential} and so is bounded to a finite range.  For the linearized model Eq.~\ref{eq:lang_delta_lin}, the potential well is parabolic so $p(\delta)$ is Gaussian with variance $\sigma_{\delta}^2 = \tau_{\delta}D_{\delta}$ (see Sect~\ref{sect:prob} for  a derivation).  The long-term variance of the change in $\delta$ is given by $\lim_{dt\rightarrow\infty} \langle [\delta(dt)-\delta(0)]^2\rangle = \lim_{dt\rightarrow\infty} \langle [\delta(dt)]^2 - 2\delta(dt)\delta(0) + [\delta(0)]^2\rangle$.  Since $\delta(dt)$ and $\delta(0)$ are uncorrelated for large $dt$, but each have variance $\sigma_{\delta}^2$, it follows for large $dt$ that  $\langle [d\delta(dt)]^2\rangle = 2\sigma_{\delta}^2 = 2\tau_{\delta}D_{\delta}$.  Since $D_{\delta}$ is measured from the slope of $\langle [d\delta(dt)]^2\rangle$ for intermediate $dt$ in Fig.~\ref{fig:diff_delta}, one can obtain $\tau_{\delta}$  from the ratio of measured values to get $\tau_{\delta} = 47$ s.    The transition between the two scaling regimes occurs at $dt = 2\tau_{\delta}$. Thus $2\tau_{\delta}$ is the time scale over which the amplitude retains some correlation.  The parameters determined here and below for $R = 1.1\times 10^{10}$ are summarized in Table~\ref{tab:parameters}. The dependences of the measured diffusivities and time scales on $R$ are shown in Sect.~\ref{sect:scaling}.

\subsubsection{Diffusion of $\dot\theta_0$}

Figure~\ref{fig:diff_dtheta} shows the mean-square change in azimuthal rotation rate $\langle(d\dot\theta)^2\rangle \equiv \langle \dot\theta_0(t+dt) - \dot\theta_0(t)\rangle$ as a function of the time interval $dt$. Here $\dot\theta_0(t) = [\theta_0(t+dt/2) - \theta_0(t-dt/2)]/dt$.  Also plotted is the same quantity derived from measurements of $\dot\theta_0(t)$  that are restricted to the range $0.9\delta_0 < \delta < 1.1\delta_0$ near the stable fixed point.  The equation $\langle d\dot\theta_0 ^2\rangle = D_{\do\theta}dt$ is fit to the latter data to obtain $D_{\dot\theta} = 2.9\times10^{-5}$ rad$^2$/s$^3$ for $R=1.1\times10^{10}$.  The diffusivity is calculated from the data with $\delta$ close to $\delta_0$ so that it can be analyzed according to the linear prediction Eq.~\ref{eq:lang_theta_lin}.  The difference between the two results is small in any case.   The plot of $\langle [d\dot\theta_0(dt)]^2\rangle$ has a plateau because the damping term causes $\dot\theta_0$ to be bounded.  Since the linear azimuthal Eq.~\ref{eq:lang_theta} also corresponds to a parabolic potential well,  $p(\dot\theta_0)$ is also Gaussian with variance $\sigma_{\dot\theta}^2 = \tau_{\dot\theta}D_{\dot\theta}/2$   which gives  $\lim_{dt\rightarrow\infty} \langle[d\dot\theta_0(dt)]^2\rangle = D_{\dot\theta}\tau_{\dot\theta}$.  Again, since $D_{\dot\theta}$ is measured from the slope of $\langle [d\dot\theta_0(dt)]^2\rangle$, one can obtain the time scale $\tau_{\dot\theta} = 6.9$ s from the ratio of measured values.  The transition between scalings occurs at $dt=\tau_{\dot\theta}$, so this  is the time scale over which $\dot\theta_0$ remains correlated.

Allowing for variations in $\delta$, for instance during cessations, the time scale corresponding to the damping term for Eq.~\ref{eq:lang_theta} is $\tau_{\dot\theta}\delta_0/\delta$, which diverges when $\delta=0$. However, such a large time scale does not exist for the LSC dynamics since cessations have a typical duration on the order of $\tau_{\delta}$  (see Sect.~\ref{sect:cess_duration}).  Since $2\tau_{\delta}$ is the correlation time for $\delta$, the damping term of the azimuthal equation may retain some autocorrelation over this time scale.   This is the longest time scale expected for the dynamics of $\dot\theta_0$, and in fact Fig.~\ref{fig:diff_dtheta} has a slightly lower plateau for $dt \stackrel{<}{_\sim} \tau_{\delta}$.  

The data in Fig.~\ref{fig:diff_dtheta} were intended to test whether the fluctuations in $\dot\theta_0$ are diffusive, but the range of the sloped region is too short to do this with  confidence. At best the diffusivity can be estimated based on the first two data points by assuming diffusive behavior.  The plateau region of Fig.~\ref{fig:diff_dtheta} corresponds to a range where the dynamics can be considered strongly damped.  For these time intervals, the acceleration term is negligible ($\ddot\theta_0\approx 0$) and the  azimuthal Eq.~\ref{eq:lang_theta} becomes 

\be
\dot\theta_0 = \frac{\delta_0\tau_{\dot\theta}}{\delta}f_{\dot\theta}(t)
\label{eq:theta_strong_damp}
\ee

\noindent  for the diffusion of $\theta_0$. It implies that fluctuations of $\theta_0$ follow a diffusive scaling $\langle d\theta_0(dt)^2\rangle = D_{\theta}dt$ where the diffusivity is

\be
D_{\theta} = \left(\frac{\delta_0\tau_{\dot\theta}}{\delta}\right)^2 D_{\dot\theta} \ .
\label{eq:diff_theta}
\ee

\noindent   Diffusive behavior for $\theta_0$ was observed in Refs.~~\cite{SXX05, XZX06, BA06a} and was used in Ref.~\cite{BA06b} to study the long-term dynamics of $\theta_0$ due to Earth's Coriolis force.  The variances $\langle(d\theta)^2\rangle$ reported in Refs. \cite{XZX06, BA06a, BA06b} do not saturate at a maximum value because $\theta_0$ represented by Eq.~\ref{eq:theta_strong_damp} has no potential terms and thus it is unbounded.  In the model of Ref.~\cite{BA06b} there are weak potential terms that would go on the right-hand-side of Eq.~\ref{eq:theta_strong_damp} due to Earth's Coriolis force which tend to align the flow in a preferred orientation and cause a net azimuthal rotation of the LSC.  However, these terms are small compared to the fluctuation size, so they do not effectively bound $\theta_0$ to a finite range \cite{BA06b}.  Since a weak Coriolis force is only relevant to long-term dynamics, a long-term  approximation can be made by assuming $\delta=\delta_0$, where the predicted value of the diffusivity of $\theta_0$ is $D_{\theta} = \tau_{\dot\theta}^2D_{\dot\theta} = 1.4\times10^{-3}$ rad$^2$/s. This can be compared with the value $D_{\theta} = 1.22\times10^{-3}$ rad$^2$/s reported in Ref.~\cite{BA06b}.  The good agreement shows that the Coriolis force model of Ref.~\cite{BA06b} is consistent with the strong-damping limit of Eq.~\ref{eq:lang_theta} near the stable fixed point.   Because the Coriolis-force terms are small they do not effect any of the short-time-scale dynamics of $\delta$ or $\dot\theta_0$ and can be ignored in the present work.  Thus the results of Ref.~\cite{BA06b} are completely consistent with and represent a long-time-scale limit of the current model.

\subsection{The probability distributions}
\label{sect:prob}

\subsubsection{The distribution $p(\delta)$}

The probability distribution $p(\delta)$ of the amplitude $\delta$ can be calculated from the steady-state solution of the Fokker-Planck equation (see, for instance, Ref.~\cite{Gi05}) which represents a balance between advection and diffusion of probability:

\be
\dot\delta_d p(\delta) = \frac{D_{\delta}}{2}\frac{d p(\delta)}{d\delta}
\ee

\noindent where $\dot\delta_d$ is the deterministic part of Eq.~\ref{eq:lang_delta}.  The solution to this differential equation is 

\be
p(\delta) \propto \exp\frac{-2V_{\delta}}{D_{\delta}}
\label{eq:prob_delta}
\ee

\noindent where the potential $V_{\delta}$ is given by Eq.~\ref{eq:potential}.  In the linear approximation, valid near the stable fixed point,  the potential $V_{\epsilon}$ from Eq.~\ref{eq:potential_lin} is parabolic, and then $p(\delta)$ is Gaussian with variance 

\be
\sigma_{\delta}^2 = \tau_{\delta}D_{\delta} \ .
\label{eq:sigma_delta}
\ee

Figure~\ref{fig:prob_delta} shows the probability distribution $p(\delta)$ derived from experimental data at $R=1.1\times10^{10}$ as open circles.  The predictions of $p(\delta)$ for both potentials $V_{\delta}$ and $V_{\epsilon}$ are plotted as well as dotted and dashed lines respectively. For the predictions the values of $D_{\delta}$ and  $\tau_{\delta}$ obtained from Fig.~\ref{fig:diff_delta} and of $\delta_0$ derived from the experimental time series for $\delta(t)$ were used.  The Gaussian shape of the peak and its variance are correctly predicted by the model.  The good match near the peaks supports the validity of the linearized model of diffusion in a potential well near the minimum.  While the nonlinear model correctly predicts a skewed distribution favoring small $\delta$, the predicted $p(\delta)$ does not match the experimental data in the tails of $p(\delta)$ very well.

For $\delta < 0.5\delta_0$, the measurements shown in Fig.~\ref{fig:prob_delta} suggest an exponential dependence of $p(\delta)$ on $\delta$. A fit of 

\be
p(\delta) =p(0)\exp\frac{B\delta}{\delta_0}
\label{eq:prob_delta_fit}
\ee

\noindent with the free parameters $B$ and $p(0)$ to data for $R=1.1\times10^{10}$ is shown in the figure.    Since cessations occur due to large flucutations in $\delta$, the statistics corresponding to cessations should be determined by the tail of $p(\delta)$ rather than by $p(\delta)$ near its peak.  Because the model is based on the assumption of the existence of the LSC, it is perhaps not surprising that it fails when the LSC is weak.   Near the end of this paper, in Sect.~\ref{sec:expanded_model}, we suggest a possible expansion of the model that could reproduce the small-$\delta$ behavior of $p(\delta)$ more accurately.

\subsubsection{The distribution $p(\dot\theta_0)$}

The peak of the probability distribution of $p(\dot\theta_0)$ can be estimated from the linearized azimuthal equation,  Eq.~\ref{eq:lang_theta_lin}.  This equation is mathematically similar to the amplitude equation Eq.~\ref{eq:lang_delta_lin}, and the steady-state solution of the Fokker-Planck equation leads to the similar result 
\be
p(\dot\theta_0) \propto \exp\left(-\frac{\dot\theta_0^2}{\tau_{\dot\theta}D_{\dot\theta}}\right)
\label{eq:prob_dtheta}
\ee

\noindent for $\dot\theta_0$ near the stable fixed point. Thus the variance of $p(\dot\theta_0)$ is

\be
\sigma_{\dot\theta}^2 = \frac{\tau_{\dot\theta}D_{\dot\theta}}{2} \ . 
\label{eq:sigma_theta}
\ee

Figure \ref{fig:prob_dtheta} shows the experimental results for $p(\dot\theta_0)$ for $R=1.1\times10^{10}$ along with the model predictions for constant damping, as well as the result of  a numerical simulation of the full model Eqs.~\ref{eq:lang_delta} and \ref{eq:lang_theta}.   The linearized model accurately describes $p(\dot\theta_0)$ as a Gaussian near the peak of the distribution, but with larger tails due to the variable damping term which allows large fluctuations in $\dot\theta_0$ when $\delta$ is small.  The simulation distribution does not match the tail of the experimental distribution very well, but does have an approximately exponential decay for large $\dot\theta_0$ like the experimental data.

\subsection{Power spectra}

To a limited extent one can test the assumption of white noise for the stochastic terms $f_{\delta}(t)$ and $f_{\dot\theta}(t)$ by examining power spectra of $\delta$ and $\dot\theta_0$.  These are shown in Fig.~\ref{fig:power_spec} for $R=1.1\times10^{10}$.  Both power spectra are mostly flat for small frequencies, as expected for white noise, but they show a rolloff for large frequencies.  The measured power spectra are a result not only of the stochastic terms but also of the response to them contained in the dynamical equations Eqs.~\ref{eq:lang_delta} and \ref{eq:lang_theta}.  Since the system spends most of its time near the stable fixed point, the amplitude response can be calculated to a good approximation from the linearized equation Eq.~\ref{eq:lang_delta_lin}.  For a single driving frequency $f$, the stochastic term can be represented by $f_{\delta}(t) = \sqrt{D_{\delta}/dt}\exp(i 2\pi f t)$.  Assuming responses of the form $\dot\delta = C\exp(i 2\pi f t + i\Phi)$, substituting these into Eq.~\ref{eq:lang_delta_lin}, and taking the magnitude of the complex solutions leads to a power $P_{\delta}$ given by the Lorentzian function 

\be
P_{\delta} \equiv |C|^2 dt = \frac{D_{\delta}}{4\pi^2 f^2 + 1/(4\tau_{\delta}^2)} \ .
\label{eq:power_delta}
\ee

\noindent  A fit of this function to the data very near the scaling crossover  gives $\tau_{\delta} = 34$ s and $D_{\delta} = 1.8\times10^{-4}$ K$^2$/s. Considering the approximations made in deriving the model, these values are in satisfactory agreement with those obtained from fits of $\langle [d\delta(dt)]^2\rangle$ (see Table~\ref{tab:parameters}).   The power spectrum $P_{\delta}$ has unexplained small features (which also appear in other data sets) that deviate from the expected function. In addition,  the rolloff appears to have an exponent somewhat more negative than $-2$, indicating that there are more low-frequency and less high-freqeuncy fluctuations than expected. Nonetheless, the shape of $P_{\delta}$ is at least roughly similar to the model prediction.  

Similarly, the response to the linearized azimuthal equation Eq.~\ref{eq:lang_theta} is 

\be
P_{\dot\theta} = \frac{D_{\dot\theta}}{4\pi^2 f^2 + 1/\tau_{\dot\theta}^2}\ .
\label{eq:power_dtheta}
\ee

\noindent A fit of $P_{\dot\theta}$ to the power spectrum of $\dot\theta_0$ gives $\tau_{\dot\theta} = 5.4$ s and $D_{\dot\theta} = 1.4\times10^{-4}$ rad$^2$/s$^3$.  This fit is excellent, and as was seen for $P_\delta$ and  $\langle [d\delta(dt)]^2\rangle$, the fit parameters are also in satisfactory agreement with those found from fits of $\langle [d\dot\theta_0(dt)]^2\rangle$.  This power spectrum is consistent with the model assumption of white noise which is filtered at high frequencies due to inertial damping.

\subsection{Correlation between $\delta$ and $|\dot\theta_0|$}

So far all of the tests of the model have assumed the damping term of the azimuthal equation Eq.~\ref{eq:lang_theta} to be constant in order to make the azimuthal equation linear.   A need for a variable $\delta$ to appear in the azimuthal equation was indicated already  by measurements that showed that $|\dot\theta_0|$ and $\delta$ are negatively correlated with $\delta$ slightly leading $|\dot\theta_0|$ \cite{BNA05, BA06a}.  This is now explained qualitatively by the model; as the damping term increases with $\delta$, the magnitude of the azimuthal rotation rate decreases.    Since $\tau_{\dot\theta} \ll \tau_{\delta}$, $\dot\theta_0$ changes faster than $\delta$. Thus the distribution of $\dot\theta_0$ comes close to a temporary stationary state for a given $\delta$ after a time of the order of $\tau_{\dot\theta}$.  This is the lead time measured in Ref. \cite{BNA05}.  It was found experimentally in Ref.~\cite{BNA05} that the delay time was about 6\% of the turnover time $\mathcal{T}$.  Using Eq.~\ref{eq:tau_theta} and $R_e = 2L^2/(\mathcal{T}\nu)$ one can estimate roughly that  $\tau_{\dot\theta} = \mathcal{T}/4$.    The measured $\tau_{\dot\theta}$ is somewhat smaller, and it has a somewhat different dependence on $R$ (see Fig.~\ref{fig:tau_R} below).  In either case there is  order-of-magnitude agreement with the measured lead time.  Numerical simulations indicate that the peak of the correlation occurs at a time of the order of $\tau_{\dot\theta}$, but that it also depends on other model parameters.

\section{The Rayleigh-number dependence of the parameters}
\label{sect:scaling}

\subsection{The time scales $\tau_{\delta}$ and $\tau_{\dot\theta}$}

The time scales $\tau_{\delta}$ and $\tau_{\dot\theta}$ predicted by the model were inferred from the mean-square variable change over time, and from power spectra, as discussed above.  The azimuthal time scale can also be estimated in the strong-damping limit from the ratio of measured diffusivities $\tau_{\dot\theta} = \sqrt{D_{\theta}/D_{\dot\theta} }$.   Each of these is non-dimensionalized in the same way as the turnover time to obtain a Reynolds number.  This gives $R_e^{\delta} \equiv L^2/(\tau_{\delta}\nu)$ which is plotted in Fig.~\ref{fig:tau_R}a and $R_e^{\dot\theta} \equiv L^2/(\tau_{\dot\theta}\nu)$ which is plotted in Fig~\ref{fig:tau_R}b.  The different methods of measuring the time scales all agree within about a factor of two.  The model predictions of Eq.~\ref{eq:tau_delta} and \ref{eq:tau_theta} are also shown in the figure.  Fits of power laws to the Reynolds numbers obtained from variance measurements in the large sample gave  $R_e^{\delta} \propto R^{0.43}$ and $R_e^{\dot \theta} \propto R^{0.20}$.  These exponents do not agree with the predicted exponents of about $1/4$ from Eq.~\ref{eq:tau_delta} and about $1/2$ from Eq.~\ref{eq:tau_theta}, respectively.   The Reynolds numbers for different samples with the same $\sigma$ and $\Gamma$ do not agree at the same $R$, indicating that the $L$-scaling of the model prediction is incorrect.  The Reynolds number $R_e = 2L^2/(\mathcal{T}\nu)$ corresponding to the plume circulation period  $\mathcal{T}$ was reported  in Ref. \cite{BFA07} to be $R_e = 0.0345R^{1/2}$ and is shown in Fig.~\ref{fig:tau_R}a.  It is seen to be very close to $R_e^{\delta}$ for the large-sample auto-correlation measurements, which implies $\tau_{\delta} \approx \mathcal{T}/2$.  The two predicted time scales were observed by several methods, and the predicted values are within an order-of-magnitude of the data, but both the $L$-scaling and the $R$-scaling of the model disagree with the measurements.

In principle, the time scales could be affected by the stochastic terms from the Langevin equations if these terms are auto-correlated over some time interval.  The time scales obained from the variances represent long-term dynamics. Thus they should be unaffected by the correlation times of any of the model terms.  A power spectrum could be modified if the corresponding stochastic term has a non-zero correlation time, which corresponds to colored noise.  Since the time scales measured from the correlation functions are close to those obtained by the mean-square variable change, the correlation times of $f_{\delta}(t)$ and $f_{\dot\theta}(t)$ must be small compared to $\tau_{\delta}$ and $\tau_{\dot\theta}$ respectively.  The correct prediction of the peaks of probability distributions based on the mean-square variable change provides support for using the timescales obtained by this method, so these measured timescales will be used as the experimental input for other model predictions.

\subsection{The amplitude $\delta_0$}

Previous work revealed correlations between the velocity of the LSC and the temperature near the side wall in the horizontal mid-plane \cite{QT02, NSSD01}.   Equation \ref{eq:delta0} predicts the  relationship between the mean temperature amplitude $\delta_0$ and the Reynolds number. It can be rearranged to get

\be
\frac{\delta_0}{\Delta T} \times \frac{R}{\sigma} \approx 18\pi (R_e)^{\frac{3}{2}} \ .
\label{eq:delta_re}
\ee

\noindent Figure \ref{fig:delta_re} shows measurements of $\delta_0/\Delta T \times R/\sigma$ vs. $R_e$ over 2.5 decades of $R$ and for $3.3 \le \sigma \le 5.5$, with data from both the medium and the large sample.  Values of $R_e$ are based on the plume circulation period determined in Ref. \cite{BFA07} from measurements with the same apparatus as the current experiments.  The Prandtl number $\sigma$ was varied by changing the mean temperature of the fluid.  The solid line shows a power-law fit with the exponent $3/2$ to the data with a free coefficient $c$ defined by $\delta_0/\Delta T \times R/\sigma = c(R_e)^{3/2}$.  The fit yields $c=159$, a factor of 2.8 larger than the prediced $18\pi$.  This power law fits  the data  within better than 20\% over 2.5 decades of $R$, and thus strongly supports the predicted relationship between $\delta_0$ and $R_e$.

\subsection{The diffusivities}

The values of the non-dimensionalized diffusivity  $D_{\delta} \times (L^2/ \Delta T^2\nu)$ are shown as a function of $R$ in Fig.~\ref{fig:diffusivity_delta}.  The non-dimensionalization  used there leads to a disagreement between data from the two samples for the same control parameters $R$ and $\sigma$,  which is undesirable.  A power law was fit to the large-sample data and yielded an exponent of -0.04. Since $(\Delta T)^2 \propto R^2$, this gives $D_{\delta} \propto R^{1.96}$.

The value of the non-dimensionalized diffusivity  $D_{\dot\theta}\times(L^2/\nu)^3$ is shown as a function of $R$ in Fig.~\ref{fig:diffusivity_dtheta}.  This non-dimensionalization is also seen to lead to disagreement between data from the two samples at the same $R$.  A power law was fit to the large-sample data to obtain $D_{\dot\theta} \propto R^{0.76}$.

\subsection{Non-dimensional parameters}

An alternate non-dimensionalization can be made by combining the three parameters from the equation for the temperature amplitude $\delta$, Eq.~\ref{eq:lang_delta}, into $\gamma \equiv (D_{\delta}\tau_{\delta})/\delta_0^2$, and the two parameters from the azimuthal equation, Eq.~\ref{eq:lang_theta}, into $D_{\dot\theta}\tau_{\dot\theta}^3$.  These are shown in Figs.~\ref{fig:diff_delta_R} and \ref{fig:diff_dtheta_R}, respectively, for various $R$ in both samples.  While there is still some disagreement for $\gamma$ between the two samples, it is smaller than in Fig.~\ref{fig:diffusivity_delta} and it is possible that this may be due to a non-Boussinesq effect \cite{ABFFGL06}.  These two dimensionless parameters have only a weak $R$-dependence.  A fit of a power law  to $D_{\dot\theta}\tau_{\dot\theta}^3$ gives an exponent of $0.14\pm 0.01$. In the large sample $\gamma$ is essentially constant, while in the medium sample and for $R \alt 4\times 10^9$ it varies as $R^{0.32}$ as shown by the solid line in Fig.~\ref{fig:diff_delta_R}.  These two non-dimensional parameters completely determine the parameter space of the linearized model.  Taking into account the variable damping term requires the additional non-dimensional parameter $\tau_{\dot\theta}/\tau_{\delta}$ which is proportional to $R^{0.23}$ in the large sample.  Since all of these non-dimensional parameters vary only weakly with $R$, the large experimental range of $R$ covers only a small region of the parameter space . It would be very interesting to learn how the parameters vary with the aspect ratio $\Gamma$ of the sample.

\section{Cessation results}
\label{sect:cessations}

\subsection{Empirical potential}

Cessations occur when $\delta$ becomes small, so the approximations near the stable fixed point may not be adequate for describing them.  It was seen in Fig.~\ref{fig:prob_delta} that the model does not accurately predict the tails of $p(\delta)$, so Eq.~\ref{eq:lang_delta} will not accurately describe the statistics of cessations, even though cessations do occur for that dynamical equation at a rate which exceeds the experimental observations by only a factor of two or three.  The problem can be seen directly in a comparison of the structure of  the Langevin equation for $\delta$ with experimental measurements. It was observed that the average of $\dot\delta$ is independent of $\delta$ both before and after a cessation when $\delta$ is sufficiently small, roughly when $\delta < \delta_0/2$. \cite{BA06a}   For small $\delta$, the model Eq.~\ref{eq:lang_delta} predicts $\dot\delta \propto \delta$, in disagreement with the data.  In this section it will be shown that this inconsistency can be repaired by using an empirical potential inferred from the experimentally measured small-$\delta$ tails of $p(\delta)$, grafted onto the parabolic potential of the model near  $\delta = \delta_0$.  

On its own, $p(\delta)$ determines the product of the frequency and of the duration of cessations.  These two values can be predicted separately for diffusion in this empirical potential well $V_{\delta,e}$ by also using the model parameters that were obtained in Sect.~\ref{sect:parameters}.
The empirical potential is obtained by fitting  Eq.~\ref{eq:prob_delta_fit} to the experimental results for $p(\delta)$ for small $\delta$.  For small $\delta$ the form of Eq.~\ref{eq:prob_delta_fit} and the Fokker-Planck result  Eq.~\ref{eq:prob_delta} then  imply 

\be
V_{\delta,e} = -\frac{BD_{\delta}\delta}{2\delta_0} + \mbox{ constant}\ . 
\label{eq:potential_cess}
\ee

\noindent  We note that this potential does not yield a fixed point at $\delta = 0$ because its derivative is finite. Nonetheless it yields a well defined potential barrier that can be used to predict properties of cessations. The time-averaged $\dot\delta$ is directly related to the potential by the equation $\dot\delta = -dV_{\delta}/d\delta$.  For small $\delta$, Eq.~\ref{eq:potential_cess} implies $\dot\delta = BD_{\delta}/(2\delta_0)$ which is independend of $\delta$ in contrast to the model Eq.~\ref{eq:lang_delta}.  The potential barrier $\Delta V_{\delta,e} = V_{\delta,e}(0)-V_{\delta,e}(\delta_0)$ can be expressed in terms of the measured $p(\delta)$ using Eq.~\ref{eq:prob_delta} to get 

\be
\frac{2\Delta V_{\delta,e}}{D_{\delta}} =\ln p(\delta_0) - \ln p(0) \equiv \Delta \ln p\ .
\label{eq:barrier_cess}
\ee

\noindent  The values of $B\gamma$ and $\gamma\Delta\ln p$ obtained from fits of $p(\delta) = p(0)\exp(B\delta/\delta_0)$ at various $R$ are shown in  Fig.~\ref{fig:prob_delta_fit}.  The peak value $p(\delta_0) = (2\pi D_{\delta}\tau_{\delta} )^{-1/2}$ is taken from the Gaussian approximation of $p(\delta)$.  Both values are seen to be nearly independent of $R$ when scaled by $\gamma$.  The inferred potential is shown in Fig.~\ref{fig:potential} for $\delta < 0.5\delta_0$ using the typical measured values $B=8.6$ and $\gamma = 0.069$, and $\Delta \ln p = 6.2$ that are a good approximation for all of the large-sample data.  The potential barrier is larger for the empirical potential then it is for the original model potential, and thus it yields a smaller cessation frequency.   

For the potential to be smooth and thus for $\dot\delta_d$ to be well defined in the corresponding Langevin equation, the small-$\delta$ limit given by $V_{\delta,e}$ and the potential $V_\delta$ near $\delta = \delta_0$ must match up at some intermediate value $\delta =  C\delta_0$.  Based on the measured $p(\delta)$, $C \approx 0.5$.  These requirements fix the parameters $B$ and $\Delta \ln p$ to be  $B = (1-C)/\gamma$ and $\Delta \ln p = (1-C^2)/(2\gamma)$.  Thus the observation that  $B\gamma$ and $\gamma\Delta \ln p$ are roughly constant is equivalent to $C$ being roughly constant, and that $p(\delta)$ has the same shape for different control parameters.  Since we do not have data covering a large range of $\gamma$, it cannot be confirmed that both $B$ and $\Delta \ln p$ scale as $1/\gamma$, but the fact that the significant decrease in $\gamma$ for small $R$ shown in Fig.~\ref{fig:diff_delta_R} does not appear in Fig.~\ref{fig:prob_delta_fit} is consistent with that conclusion over a small range of $\gamma$.

\subsection{The rate of cessations}

The equation for $\delta$ does not contain an inertia term, and corresponds to  pure diffusion in a potential well. Thus, if the root-mean-square amplitude of this diffusive motion is small compared to its mean value $\delta_0$, then successive cessations will follow Poissonian statistics. This will be the case when the diffusivity $D_\delta$ is small compared to the depth of the well, and this condition is satisfied for the physically relevant parameter values.  

The experimental results for the time-averaged frequencies of cessations $\omega_c$, measured in events per unit time, are not very accurate because cessations occurred only about once or twice per day. Nonetheless a comparison between experiment and the model is useful. The rate of cessations can be calculated using the model of diffusion across a potential barrier, which is analagous to the Arrhenius-Kramers problem \cite{Kr40}.  Stochastic fluctuations with strength $D_{\delta}$ drive the amplitude $\delta$ in a potential well.  A cessation occurs when $\delta$ fluctuates from the bottom of the potential well at $\delta = \delta_0$ to the top at $\delta = 0$. Thus, for a cessation to occur, fluctuations must overcome a barrier $\Delta V \equiv V(0) - V(\delta_0)$.  First we shall use the empirical potential barrier $\Delta V_{\delta,e}$. It can be obtained  from Eq.~\ref{eq:barrier_cess} and the data in Figs.~\ref{fig:diff_delta_R} and \ref{fig:prob_delta_fit}. One sees that the ratio of the potential barrier to the diffusivity given by $2\Delta V_{\delta,e}/D_{\delta}$ is around 6 or larger.  This means that the Arrhenius-Kramers equation for the large-barrier approximation \cite{Kr40} can be applied here.  In that approximation the rate of escape is proportional to $\exp(-2\Delta V_{\delta,e}/D_{\delta})$.  The prefactor depends on expansions of integrals around the peak and minimum of $p(\delta)$ with the result depending on the curvature of the potential at these points.  The full solution is $\omega_c = (B/\tau_{\delta})\sqrt{\gamma/(2\pi)}\exp(-2\Delta V_{\delta,e}/D_{\delta})$. While the exponential dependence always has the same form in the large-barrier limit, the proportionality to $B\sqrt{\gamma}$ is a result of the shape of the potential near $\delta=0$.  This prediction for $\omega_c$ is shown in Fig.~\ref{fig:barrier_rate} as triangles, along with experimental data from Ref.~\cite{BA06a} which are given as circles.   The prediction is seen to be in agreement with the experimental data within about a factor of 2.  The cessation rate is nearly independent of $R$ for the large sample, and decreases for the medium sample with decreasing $R$.    This plot roughly follows the same trend as $\gamma$, see Fig.~\ref{fig:diff_delta_R}, confirming that $\gamma$ is the most relevant parameter for determining the rate of cessations. This it must be since it is the only dimensionless parameter in the relevant Langevin equation.

For comparison, the prediction in the large-barrier limit for the model potential Eq.~\ref{eq:potential} is $ \omega_c = \exp(-2\Delta V_{\delta}/D_{\delta})/(2\pi\tau_{\delta})$,  but the large-barrier limit is not as good an approximation for this model in the experimental parameter range because $\Delta V_{\delta}$ is too small  (see the potential barriers in Fig.~\ref{fig:potential}). For instance, for $R = 1.1\times 10^{10}$ it yields about 14 cessations per day, which is an order of magnitude larger than the experiment. However, when the complete model, Eqs.~\ref{eq:lang_delta} and \ref{eq:lang_theta}, is integrated numerically for the same $R$ instead of using the large-barrier analytic expression, one finds about 3.8 cessations per day \cite{BA07a}, which is only a factor of two larger than the experiment. 
It turns out that the prediction based on the empirical $p(\delta)$ and the prediction based on the parabolic potential of the linearized model give about the same result for $\omega_c$, but the functional dependence on $\gamma$ differs between the two due to the different shape of the potentials around $\delta=0$.

\subsection{The duration of cessations}
\label{sect:cess_duration}

For $\delta  \alt 0.5\delta_0$ it was observed experimentally \cite{BA06a} that during cessations the average over all events of the  magnitude $ \langle | \dot\delta | \rangle$ of the rate of change   of  $\delta$ was independent of $\delta$. Thus the amplitude drop and recovery are symmetric and  follow the equation 
\be
\delta(t) = \delta_m + \langle|\dot\delta|\rangle |t|
\label{eq:cess_amp_rate}
\ee
\noindent where $t$ is the time elapsed since the cessation. We calculate the average duration of a cessation to be $\langle\Delta t\rangle = (\delta_0-2\delta_m)/\langle|\dot\delta|\rangle$. Here $\delta_m = 0.095\delta_0$ is the average minimum measured amplitude during cessations, which is just slightly larger than zero in any experimental measurement.    The value of $\langle \dot\delta(\delta)\rangle $ for the rise can be calculated from the experimentally obtained small-$\delta$ tail of $p(\delta)$ to be $\langle\dot\delta\rangle = -dV_{\delta,e}/d\delta = BD_{\delta}/(2\delta_0)$ using the empirical $V_{\delta,e}$ from Eq.~\ref{eq:potential_cess}.  This $\langle\dot\delta\rangle$ is independent of $\delta$ in agreement with experiment \cite{BA06a}. The average duration of cessations is thus predicted to be $\langle\Delta t\rangle \approx (\delta_0-2\delta_m)/[BD_{\delta}/(2\delta_0)]$.  The average measured cessation duration, as well as the predicted value based on the empirical potential, are shown in Fig.~\ref{fig:cess_dur} for various $R$.  Both data and the prediction of $\Delta t$ are roughly proportional to $\tau_{\delta}$, but the prediction is an overestimate by about a factor of 2.  While the amplitude behavior during cessations is inconsistent with the proposed model Eq.~\ref{eq:lang_delta} in the sense that it is responsible for the tails of $p(\delta)$, using the tails of the measured $p(\delta)$ as experimental input into the diffusion model allowed the prediction that $\langle\dot\delta\rangle$ during cessations is constant and approximately proportional to $\delta_0/\tau_{\delta}$.

\subsection{Angular change during cessations}
\label{sect:cess_delta_theta}

Within experimental resolution the net angular change $\Delta \theta$ during cessations was found to have a uniform probability distribution $p(\Delta\theta)$ \cite{BNA05}. Numerical simulations based on the model equations Eqs.~\ref{eq:lang_delta} and \ref{eq:lang_theta} confirmed this within their resolution \cite{BA07a}.  A viable but ad hoc explanation of this result was that, once the LSC ceases, there is no memory of its original orientation and the re-organization of the new circulation will occur in an arbitrary new orientation. Here we offer an alternative view of this phenomenon. We present arguments showing that a near-uniform distribution can be due to large azimuthal fluctuations that occur when the inertial damping term in Eq.~\ref{eq:lang_theta} becomes small. This in occurs only when $\delta$ becomes small during cessations.  

Equation \ref{eq:prob_dtheta} predicts that, over a given time period $\delta t$, the orientation diffuses, yielding a Gaussian distributed $p(\Delta\theta)$. The typical orientation change due to diffusion near the fixed-point amplitude $\delta_0$ is $d\theta_{rms} =  \sqrt{D_{\theta} \Delta t}$, 
which with Eq.~\ref{eq:lang_theta} yields $d\theta_{rms}  \approx \tau_{\dot\theta}\sqrt{D_{\dot\theta}\Delta t}$. For the average cessation period $\langle\Delta t\rangle \approx 1.34 \mathcal{T}$ \cite{BA06a} one then has $d\theta_{rms} \approx  0.3$ rad. 
Because the inertial damping decreases as $\delta$ decreases during a cessation (see Eq.~\ref{eq:lang_theta}), the angular change is typically larger during cessations then it is when $\delta$ is near $\delta_0$ or larger.  In the strong-damping approximation, the diffusivity of $\theta_0$ is given by Eq.~\ref{eq:diff_theta} and is larger at smaller $\delta$, and the time dependence of $\delta(t)$ is given by Eq.~\ref{eq:cess_amp_rate}.  The mean-square change in amplitude, when  integrated over the duration of a cessation, gives 
\be
\langle [d\theta_0(\Delta t)]^2\rangle = \int_{-\Delta t/2}^{\Delta t/2} D_{\theta}[\delta(t)] dt = 2D_{\dot\theta}\tau_{\dot\theta}^2\delta_0 \Delta t/\delta_m\ .
\ee
Thus the typical angular change during cessations $\sigma_{\Delta\theta} \equiv \sqrt{\langle [d\theta_0(\Delta t)]^2\rangle} \approx 1.4$ rad at $R=1.1\times10^{10}$.  This is much larger than the angular change for constant damping with $\delta$ near $\delta_0$ that was illustrated above.

There is an additional contribution during cessations when the flow reverses -- that is when $\delta$ becomes less than zero.  This can occur due to a continuous meandering of $\delta$ below zero, but in our analysis the absolute value $|\delta|$ is taken and $\pm\pi$ added  alternatively to $\theta_0$ whenever a reversal occurs.  For an odd number of reversals, there is a net contribution of $\pi$ to $\Delta\theta$. The percentage of cessations with reversals depends on the potential barrier to cessations and how the threshold for cessations is defined.  This would be 50\% if a cessation was counted only if $\delta$ crossed zero, but because the LSC cannot be resolved experimentally when $\delta \approx 0$, a cessation is counted when $\delta$ drops below $0.15\delta_0$.  The chance of getting an odd number of reversals in a cessation can be estimated in the large-barrier limit to be $A = 0.5\exp[2[V_{\delta,e}(0) - V_{\delta,e}(0.15\delta_0)]/D_{\delta}] \approx 0.5\exp(-0.15B) \approx 0.14$ at $R=1.1\times10^{10}$.   While this is at best an order-of-magnitude estimate, numerical simulations indicate that there are an odd number of reversals for a fraction $A=0.32$ of cessations at this value of $R$.  The combination of diffusion and reversals results in a double-peaked distribution at 0 and $\pi$ for small azimuthal fluctuations, but with larger fluctuations, both peaks spread out.  This predicted distribution is reduced to the range $0~ \mathrm{to} ~\pi$ by the transformation $\Delta\theta_{red} = \pi - |\pi-|\Delta\theta~\mathrm{mod}~2\pi ||$ so that $\Delta\theta_{red}$ is the smaller of the choices of angular change in either direction during the cessation.  This transformation is made because of the non-uniqueness of $\theta_0$ in which a change of $\Delta\theta$ and $\Delta\theta \pm 2\pi$ are indistinguishable unless the orientation can be smoothly traced in time, which cannot practically be done during cessations.  The distribution can be expressed as

\begin{eqnarray}
p(\Delta\theta_{red}) &\propto& \sum_{n=-\infty}^{\infty} (1-A)\exp\left[-\frac{(\Delta\theta+2\pi n)^2}{2\sigma_{\Delta\theta}^2}\right] \nonumber \\
& +& \sum_{n=-\infty}^{\infty} A\exp\left[-\frac{(\Delta\theta-\pi+2\pi n)^2}{2\sigma_{\Delta\theta}^2}\right]. 
\label{eq:prob_delta_theta}
\end{eqnarray}

\noindent If both peaks are assumed to have the same height, the net distribution is within 10\% of uniform if $\sigma_{\Delta\theta}  > 1.35$ rad.    In the limiting case where $A=0$ (no reversals), for $p(\Delta\theta)$ to be within 10\% of uniform would require $\sigma_{\Delta\theta} > 2.70$ rad.  For stronger diffusion, the absolute angular change can increase, but $p(\Delta\theta_{red})$ remains nearly uniform.  Experimental data from Ref.~\cite{BA06a} for $10^{9} < R < 1.1\times10^{10}$ are plotted in Fig.~\ref{fig:prob_delta_theta}, along with the predictions of Eq.~\ref{eq:prob_delta_theta} for $R=1.1\times10^{10}$ and $R=10^{9}$ using $A=0.32$.  The model result is consistent with the data without any fit parameters. The predictions for both ends of the $R$ range, as well as the uniform distribution, are consistent with the data because the experimental probability distribution has fairly large error bars since cessations are so rare that only a few hundred have been measured.

Experiments reported in Ref.~\cite{XX07} with $\Gamma = 1/2$ for $1.5\times 10^{10} \le R \le 7.2\times 10^{10}$ found $p(\Delta\theta)$ peaked at 0 and $\pi$ with $p(\pi/2) \approx 0.3p(0)$.  Although the effect of changing $\Gamma$ on the model parameters is unknown, a single LSC roll was observed, so the model should apply to those experiments with different parameters.   The measured $p(\Delta\theta)$ is consistent with the predicted functional form of the sum of Gaussians, one centered at each integer multiple of $\pi$.

\section{An expanded model for small $\delta$}
\label{sec:expanded_model}

Measurements of $p(\delta)$ and $\dot\delta$ for small $\delta$ indicate that the Langevin equation for $\delta$ should be dominated by a constant term for $\delta \stackrel{<}{_\sim} 0.5\delta_0$, in contrast to the model equation Eq.~\ref{eq:lang_delta} which has $\dot \delta \sim \delta$ when $\delta$ is small.  In this section we outline how the model might be expanded so as to account for the small-$\delta$ behavior, but leave the details of this expansion to future work.

The expansion consists of the inclusion of  the heat-transport equation 
\be
\dot T + \vec u\cdot \vec\nabla T = \kappa \nabla^2 T
\label{eq:Tdot}
\ee
 to describe the top and bottom boundary-layer heat-conduction.  When the LSC is weak, $\delta$ and $U$ are small and the advective term of Eq.~\ref{eq:Tdot}  can be neglected.  The diffusion term contributes only in the thermal boundary layers and can be estimated as $\kappa\nabla^2 T \approx \Delta T/(2l^2)$, where $l$ is the width of one boundary layer. It is given approximately by $l = L/(2\mathcal{N})$ \cite{GL02}. This term contributes in the fractional volume $2l/L$.  Using Eq.~\ref{eq:temp_profile}, the volume averaged temperature change is  $\dot T = 4\dot\delta/(3\pi)$.  Combining these terms gives $\dot\delta = 3\pi\kappa\Delta T\mathcal{N}/(2L^2) \approx 0.03$ K/s for $R=1.1\times10^{10}$.  This is a constant, i.e. independent of $\delta$,  as required by the data for $p(\delta)$. It is  within an order-of-magnitude of the experimental value $\dot\delta \approx 1.34 \delta_0/\mathcal{T}\approx 0.007$ K/s reported in Ref. \cite{BA06a}.  Using the result $\mathcal{N} \propto R^{1/3}$ for the large sample \cite{NBFA05}, the prediction gives $\dot\delta \propto R^{4/3}$.  This is close to the measured $\dot\delta \propto \delta_0/\mathcal{T} \propto R^{5/4}$ reported in Ref.  \cite{BA06a}.  These values are consistent with the conclusion that the dominating driving term for small $\delta$ is due to thermal diffusion across the boundary layers.  At least for small $\delta$, this replaces the need to use the momentum equation and the assumption $\delta \propto U$.  
 
 A more complete model that applies for all ranges of $\delta$ might consider the LSC velocity $U$ separately from $\delta$.  This separation would be more realistic since $\delta$ and $U$ fluctuate separately, even though the model produced cessations and rotations without taking this into account.  Such a model would likely include the momentum Eqs.~\ref{eq:uave} and \ref{eq:theta_ave} as well as the heat transport equation, for a total of three differential equations for the three parameters $U$, $\theta$, and $\delta$.  The advective term of the heat-transport equation couples it to $U$, while the buoyancy term of Eq.~\ref{eq:nav} for the LSC velocity couples it to $\delta$.   
The fact that only one equation was necessary to describe the LSC strength near the stable fixed point suggests that the time scale for the coupling of $\delta$ and $U$ is short compared to $\tau_{\delta}$, so that the dynamics of $\delta$ and $U$ can to lowest order be described by a single equation.

\section{Summary and Conclusions}

We demonstrated that many aspects of the dynamics of the LSC can be described by two coupled stochastic ODEs that are motivated by the Navier-Stokes equations. One of the equations is for the amplitude of the azimuthal temperature variation $\delta$ that is induced by the circulation, and the other is for the azimuthal orientation $\theta_0$ of the near-vertical LSC circulation plane. The $\delta$-equation represents the balance between the driving due to buoyancy and the dissipation due to the drag across viscous boundary layers near the walls. The $\theta_0$ equation is coupled to the $\delta$-equation by an advective term that represents the angular momentum of the circulation. Both equations are driven by a Gaussian white noise term that represents the action of the small-scale turbulent fluctuations on the large-scale circulation.  

The original definition of reorientations as a relatively large and fast angular change was cumbersome because there was no clear distinction between small fluctuations and large reorientations \cite{BNA05}.  Now all of the azimuthal dynamics -- including meandering, larger and faster rotations, and the large azimuthal changes associated with cessations -- are simply described as diffusive fluctuations in a potential well $V_{\dot\theta}$ given by Eq.~\ref{eq:potential_theta} and with a  curvature that depends on $\delta$.  Larger rotations tend to occur when $\delta$, and thus the well curvature, is smaller, and the large angular change during cessations (with $\delta$ near zero) occurs because the potential well becomes nearly flat so that the azimuthal motion is almost unconfined.  Reversals of the LSC with $\Delta\theta \simeq \pi$ are not distinct events in this description.  Crossings of the potential maximum at $\delta=0$ tend not to be reversals because this is when the azimuthal fluctuations are at their largest.  

Previously, we had suggested that the uniform distrubution of $p(\Delta\theta)$ for cessations occurs because the LSC loses any memory of its orientation during cessations \cite{BA06a}.  This is the expected result if the LSC completely breaks up.  In Sect.~\ref{sect:cess_delta_theta} we showed that a good approximation to this distribution can occur even if $\delta$ remains  finite during  cessations, provided $\delta$ becomes small enough so that  the azimuthal diffusion is large over the durations of cessations. 

By studying power spectra and probability distributions, it was found that the stochastic driving terms can to a good approximation be described by Gaussian white noise. The model contains two time scales:  $\tau_{\dot\theta}$ and $\tau_{\delta}$. These were measured indirectly by several methods to be within a factor of two or so of the prediction; however, the predicted dependence on the Rayleigh number $R$ and the sample height $L$ differed somewhat from the data.  The prediction for the dependence of the mean temperature amplitude $\delta_0$ of the LSC on $R$ agreed well with measurements over the large range of $R$ and the small range of the Prandtl number $\sigma$ that were explored.  

The model accurately describes the dynamics and much of the $R$-dependence using the potential near the minimum where $\delta\approx \delta_0$ and predicts the existence of cessations near $\delta = 0$.  However, it fails to predict quantitatively the behavior far from the potential minimum, including the tails of $p(\delta)$ and details about cessations.   Using the measured $p(\delta)$ as empirical experimental input to the model, the scaling behavior of the frequency, angular change, and amplitude behavior during cessations could be reproduced quite well. This supports using a model of diffusion in a potential well for the LSC even though the originally proposed model equations must be modified for small $\delta$.  In the preceding section we suggest that the modification may consist of the inclusion of the driving by the diffusive heat transport across the top and bottom thermal boundary layers which becomes important when the driving due to $\delta$ becomes small.   

In agreement with experiment, the model predicts a rate of cessations that is roughly uniform at about one to two per day over a wide range of Rayleigh numbers, from $3\times 10^8 < R< 10^{11}$ \cite{BA06a}.   On the other hand, the model of diffusion over a potential barrier  suggests a very sensitive dependence of the rate of cessations on the model parameters, at least when the potential barrier is not small compared to the diffusivity.  The lack of a dependence on $R$ can be understood because the barrier ratio $2V/D_{\delta}\propto \gamma^{-1}$ is nearly independent of $R$.  It is conspicious that the two dimensionless parameters $\gamma$ and $D_{\dot\theta}\tau_{\dot\theta}^3$ that determine the dynamics of $\delta$ and $\dot\theta_0$ respectively are both nearly constant over the 2.5 decades of $R$ that were studied.

The only known dynamical behavior of the LSC that is conspicuously missing from the model is a description of the twisting oscillation of the LSC \cite{FA04,FBA08}.   There is no signal of the twisting oscillation in measurements at the mid-height of the sample, so it seems that the oscillations are independent of the azimuthal diffusion and re-orientations described by the model.  The Nusselt number, or non-dimensional heat flux, is another important aspect of RBC not represented in the model.  The Nusselt number is understood to be controlled by the thermal boundary layers and not the LSC, so its absence from a model of the LSC which does not include these boundary layers is not surprising.

Here the LSC was studied in a container with cylindrical symmetry.  A previous paper described measurements and a model of the symmetry-breaking effect of  Earth's Coriolis force on the LSC \cite{BA06b}. That model is consistent with the strong-damping limit of the current model.  A later project will consider the effects of various asymmetries of the system on the LSC in which perturbative terms can be added to the model equations  Eqs.~\ref{eq:lang_delta} and \ref{eq:lang_theta}.

\section{Acknowledgment}

This work was supported by Grant DMR07-02111 of the US National Science Foundation.

\newpage

\begin{table}
\begin{center}
\begin{tabular}{ccccc}
& $\langle [dx(dt)]^2\rangle$	&$ P_x(\omega)$ & prediction \\
\hline
$\tau_{\delta}$ (s) & 47  & 34   & 85 \\
$\tau_{\dot\theta}$ (s)  & 6.9  & 5.4   & 13 \\
$D_{\delta}$ (K$^2$/s) & $6.4\times10^{-5}$ & $1.8\times10^{-4}$  & --\\ 
$D_{\dot\theta}$ (rad$^2$/s$^3$) & $2.9\times10^{-5}$ & $1.4\times10^{-4}$   & --\\
\end{tabular}
\caption{The time scales $\tau_{\delta}$ and $\tau_{\dot\theta}$, together with the model predictions Eqs.~\ref{eq:tau_delta} and \ref{eq:tau_theta}, and the diffusivities $D_{\delta}$ and $D_{\dot\theta}$, obtained from the mean-square change $\langle [dx(dt)]^2\rangle$ of the variables over time and from power spectra $ P_x(\omega)$ for $R=1.1\times10^{10}$.}
\label{tab:parameters}
\end{center}
\end{table}

\clearpage

\begin{figure}
\caption{ A diagram of the LSC showing the coordinate $\phi$, maximum of the velocity profile $U$, and viscous boundary-layer width $\lambda$ (not to scale).}
\label{fig:wind_phi}
\end{figure}

\begin{figure}                                                
\caption{Solid line:  The potential well $V = [-\delta^2/2 + 2\delta^{5/2}/(5\sqrt{\delta_0}) ]/\tau_{\delta}$ in which $\delta$ meanders diffusively.   Dashed line:  a parabolic potential well $V_{\epsilon} = V(\delta_0) + (\delta-\delta_0)^2/(4\tau_{\delta})$.  Dotted line:  the potential for $\delta < 0.5\delta_0$ inferred from the measured $p(\delta)$.   $\Delta V$ is the predicted potential barrier for cessations.}  
\label{fig:potential}                                        
\end{figure}

\begin{figure}                                                
\caption{The mean-square change in amplitude $\langle(d\delta)^2\rangle$ as a function of the time interval $dt$ for $R=1.1\times10^{10}$.  Solid line:  a fit of  $\langle(d\delta)^2\rangle = D_{\delta} dt$ to the data for intermediate $dt$ gives the diffusivity $D_{\delta}$.  Dotted line:  a constant $2D_{\delta}\tau_{\delta}$ fit to data with large $dt$, together with $D_\delta$,  yields $\tau_{\delta}$. }  
\label{fig:diff_delta}                                        
\end{figure}

\begin{figure}                                                
\caption{Dots: the mean-square change in azimuthal rotation rate $\langle(d\dot\theta)^2\rangle$ as a function of the time interval $dt$ for $R=1.1\times10^{10}$.  Open circles:  modified time series using only azimuthal steps $d\theta_0$ when $0.9\delta_0 < \delta < 1.1\delta_0$.  Solid line:  a fit of $\langle d\dot\theta_0 ^2\rangle = D_{\do\theta}dt$ to the open circles for $dt < 6$ s.  Dotted line:  a constant $D_{\dot\theta}\tau_{\dot\theta}$ fit to data with large $dt$. Together with $D_{\dot\theta}$ it yields  $\tau_{\dot\theta}$ .}  
\label{fig:diff_dtheta}                                        
\end{figure}

 \begin{figure}
\caption{The probability distribution of the amplitude $p(\delta)$ for $R=1.1\times10^{10}$.  Dashed line:  Gaussian fit to data.  Dotted line:  model prediction from Fokker-Planck equation.  Solid line:  an exponential fit of $p(\delta) = p(0)\exp(B\delta/\delta_0)$ to data for $\delta < 0.6\delta_0$.  }
\label{fig:prob_delta}
\end{figure}

\begin{figure}                                                
\caption{The probability distribution of the azimuthal rotation rate $p(|\dot\theta_0|)$ over a single time step  $dt\approx2.5$ s for $R=1.1\times10^{10}$.  Circles:  experimental data.  Triangles:  simulation data.  Dotted line:  the Gaussian distribution predicted for a constant damping term with $\delta = \delta_0$.   Solid line:  exponential fits to the tails of the distributions. }  
 \label{fig:prob_dtheta}                                        
\end{figure}

\begin{figure}                                                
 \caption{The power spectra $P_{\delta}$ (dotted line) and $P_{\dot\theta}$ (thin solid line) derived from the experiment.  Dashed line:  a fit of a Lorentzian near the crossover of $P_{\delta}$.   Thick solid line:  a  fit of a  Lorentzian to $P_{\dot\theta}$. }  
 \label{fig:power_spec}                                        
\end{figure}

\begin{figure}                                                
\caption{(a) The Reynolds number $R_e^{\delta} \equiv L^2/(\nu\tau_{\delta})$.  (b) The Reynolds number $R_e^{\dot\theta} \equiv L^2/(\nu\tau_{\dot\theta})$.  Each are measured by several methods for a wide range of $R$.  Circles:  from the mean-square variable change over time.  Triangles: from power spectra.  Squares:  from the strong-damping approximation of $\tau_{\dot\theta} = \sqrt{D_{\theta}/D_{\dot\theta}}$.  Solid symbols:  medium sample.  Open symbols: large sample.  Solid lines:  prediction from model Eqns.~\ref{eq:tau_delta} and \ref{eq:tau_theta}.  Dashed line:  the Reynolds number corresponding to the turnover $R_e = 0.0345R^{1/2}$ from Ref. \cite{BFA07}.  Dotted lines:  power law fits to the open circles.}  
\label{fig:tau_R}                                        
\end{figure}

 \begin{figure}
\caption{The measured value of $\langle\delta\rangle/\Delta T \times R/\sigma$ as a function of  $R_e$.  Solid circles:  medium sample, $\sigma = 4.4$.  Open circles:  large sample; $\sigma = 4.4$.  Up-pointing triangles:  medium sample, $\sigma = 5.5$.  Down-pointing triangles:  medium sample, $\sigma = 3.3$.  Solid line: a fit of the predicted power law $\langle\delta\rangle/\Delta T \times R/\sigma = c R_e^{3/2}$ to all of the data.}
\label{fig:delta_re}
\end{figure}

\begin{figure}                                                
\caption{The non-dimensionalized diffusivity $D_{\delta} \times (L^2/ \Delta T^2\nu)$ for various $R$.  Solid circles:  medium sample.  Open circles: large sample.  Solid line: a power-law fit to the large sample data. }  
\label{fig:diffusivity_delta}                                        
\end{figure}

\begin{figure}                                                
\caption{The non-dimensionalized diffusivity $D_{\dot\theta} \times(L^2/\nu)^3$ as a function of $R$.  Solid circles: medium sample.  Open circles: large sample.  Solid line:  a power-law fit of the large sample data.}  
\label{fig:diffusivity_dtheta}                                        
\end{figure}

\begin{figure}                                                
\caption{The non-dimensionalized diffusivity $\gamma = D_{\delta}\tau_{\delta}/{\delta_0^2}$ for various $R$.  Solid circles:  medium sample.  Open circles: large sample. Solid line: a power law with an exponent of 0.32. }  
\label{fig:diff_delta_R}                                        
\end{figure}

\begin{figure}                                                
\caption{The non-dimensionalized diffusivity $D_{\dot\theta}\tau_{\dot\theta}^3$ as a function of $R$.  Solid circles: medium sample.  Open circles: large sample.  Solid line:  power law fit to the data.}  
\label{fig:diff_dtheta_R}                                        
\end{figure}

 \begin{figure}
\caption{The coefficients obtained from a fit of $p(\delta) = p(0)\exp(B\delta/\delta_0)$ to the measured $p(\delta)$ for various $R$.  Circles:  $B\gamma$.  Triangles: $\gamma\Delta \ln p  \equiv \gamma \times [\ln p(\delta_0) -\ln p(0)] = 2\gamma V/D_{\delta}$.  Solid symbols: medium sample.  Open symbols: large sample.}
\label{fig:prob_delta_fit}
\end{figure} 

\begin{figure}                                                
\caption{Black circles: The measured frequency of cessations $\omega_c$ from Ref.~\cite{BA06a} for various $R$.   Triangles:  prediction using the measured tail of $p(\delta)$ in the large-barrier limit for diffusion in a potential well. Solid symbols: medium sample.  Open symbols: large sample.  }    
\label{fig:barrier_rate}                                    
\end{figure}

\begin{figure}                                                
\caption{The average duration of cessations $\langle \Delta t \rangle /\tau_{\delta}$ for various $R$.  The duration is calculated as the time that $\delta$ remains below the threshold $\delta_0/2$.  Circles: experimental data.  Triangles: predictions based on measurements of the tails of $p(\delta)$.  Solid symbols: medium sample.  Open symbols: large sample.}
\label{fig:cess_dur}                                       
\end{figure}

 \begin{figure}
\caption{The probability distribution $p(\Delta\theta_{red})$ of the orientation change during cessations, reduced to the range $0 ... \pi$  by the transformation $\Delta\theta_{red} = \pi - |\pi-|\Delta\theta~\mathrm{mod}~2\pi ||$.  Solid circles:  data from \cite{BA06a}.  Solid line:  uniform distribution.  Dotted line;  prediction from Eq.~\ref{eq:prob_delta_theta} for $R=10^9$.  Dashed line:  prediction from Eq.~\ref{eq:prob_delta_theta} for $R=1.1\times10^{10}$. }
\label{fig:prob_delta_theta}
\end{figure}

\clearpage

\begin{figure}
\centerline{\includegraphics[width=2.25in]{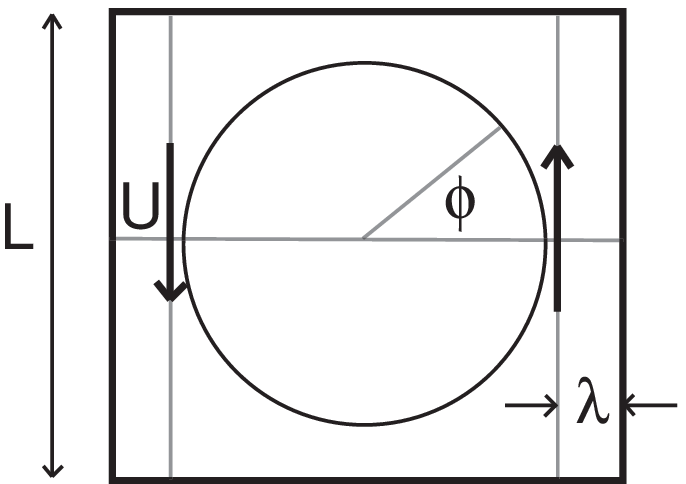}}
 \centerline{Fig. 1, Eric Brown, Physics of Fluids.}
\end{figure}

\begin{figure}                                                
\centerline{\includegraphics[width=3in]{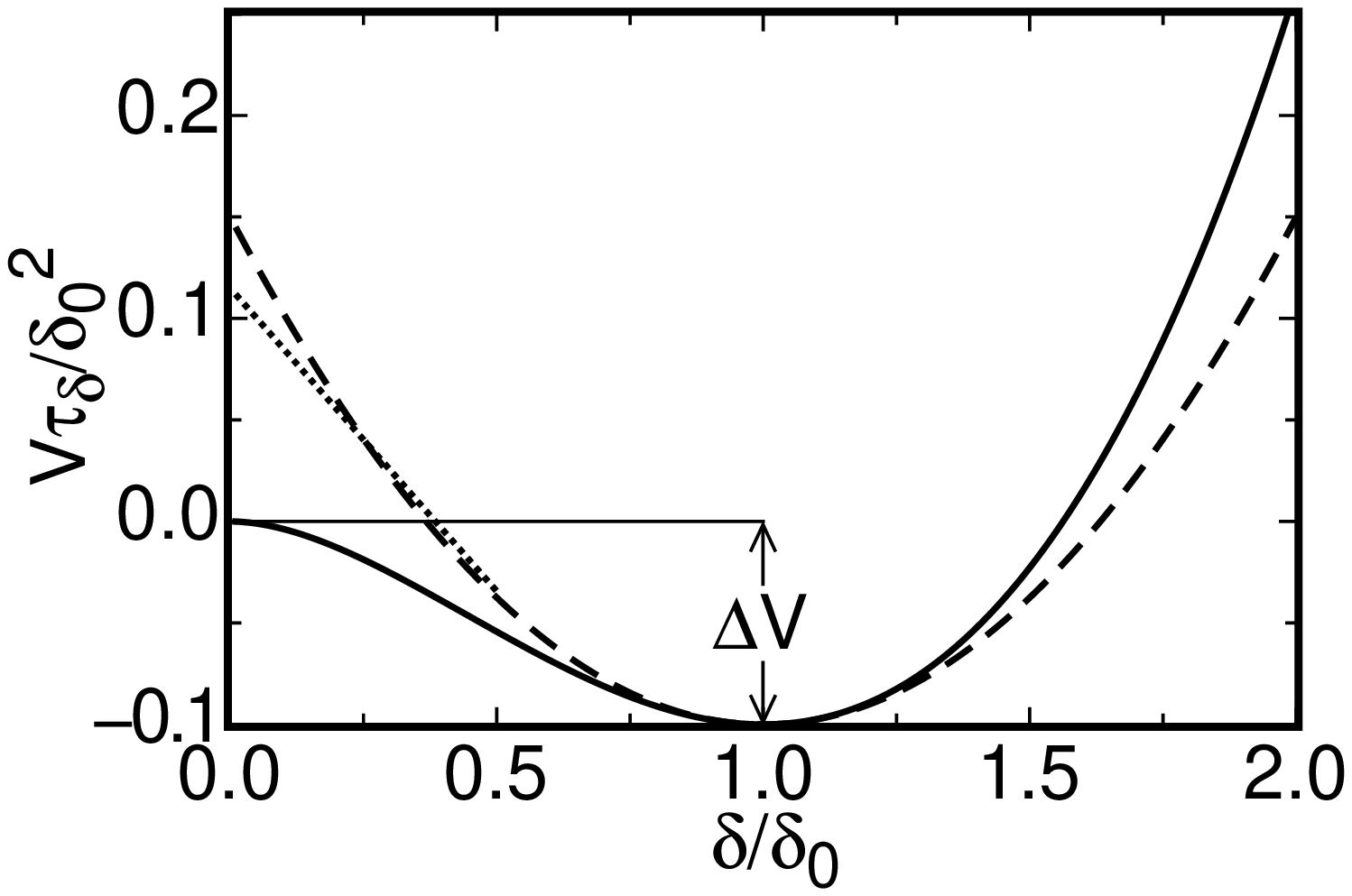}}
 \centerline{Fig. 2, Eric Brown, Physics of Fluids.}  
\end{figure}

\begin{figure}                                                
\centerline{\includegraphics[width=3.25in]{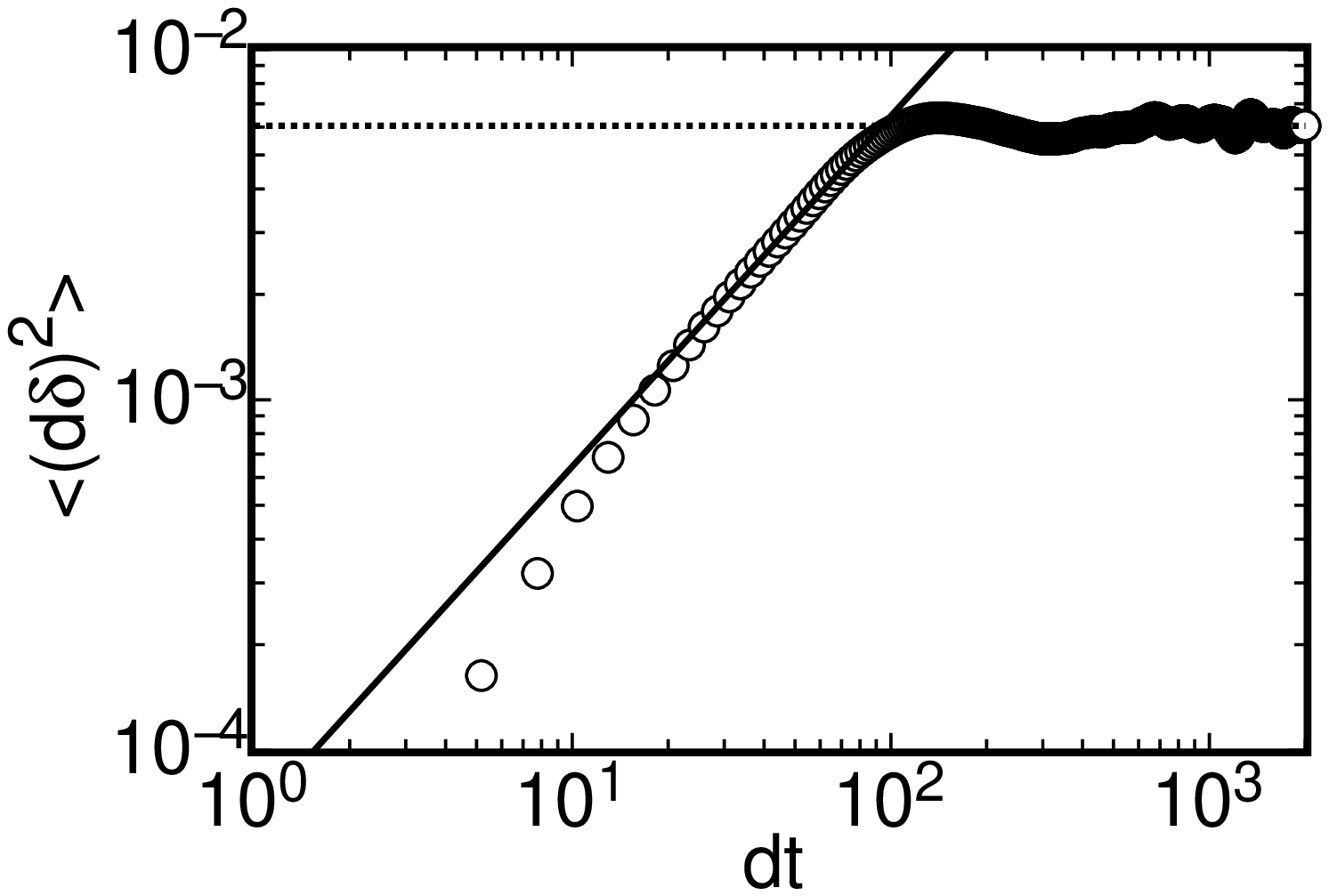}}
 \centerline{Fig. 3, Eric Brown, Physics of Fluids.}  
\end{figure}

\begin{figure}                                                
\centerline{\includegraphics[width=3.25in]{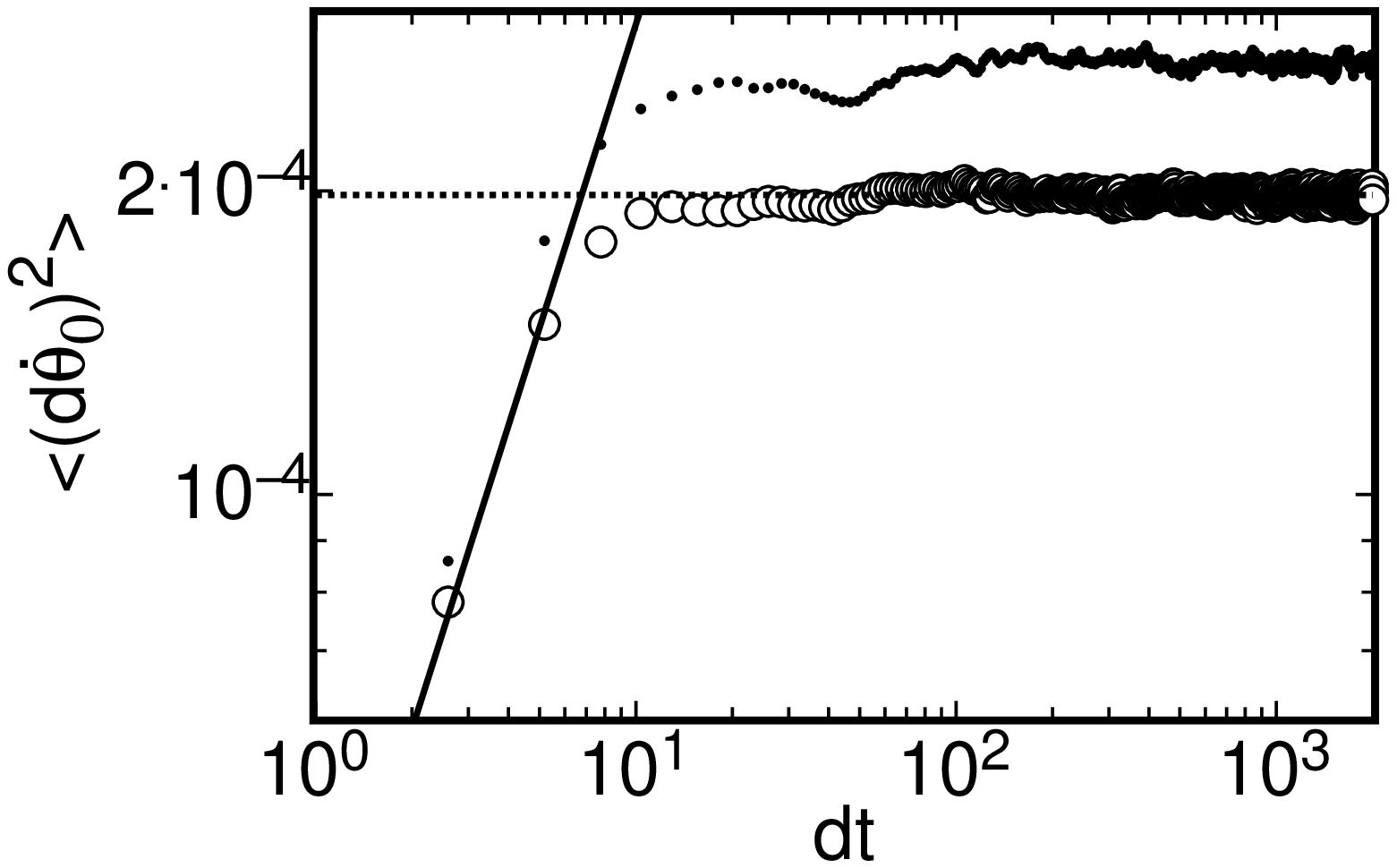}}
 \centerline{Fig. 4, Eric Brown, Physics of Fluids.}  
\end{figure}

 \begin{figure}
\center{ \includegraphics[width=3.25in]{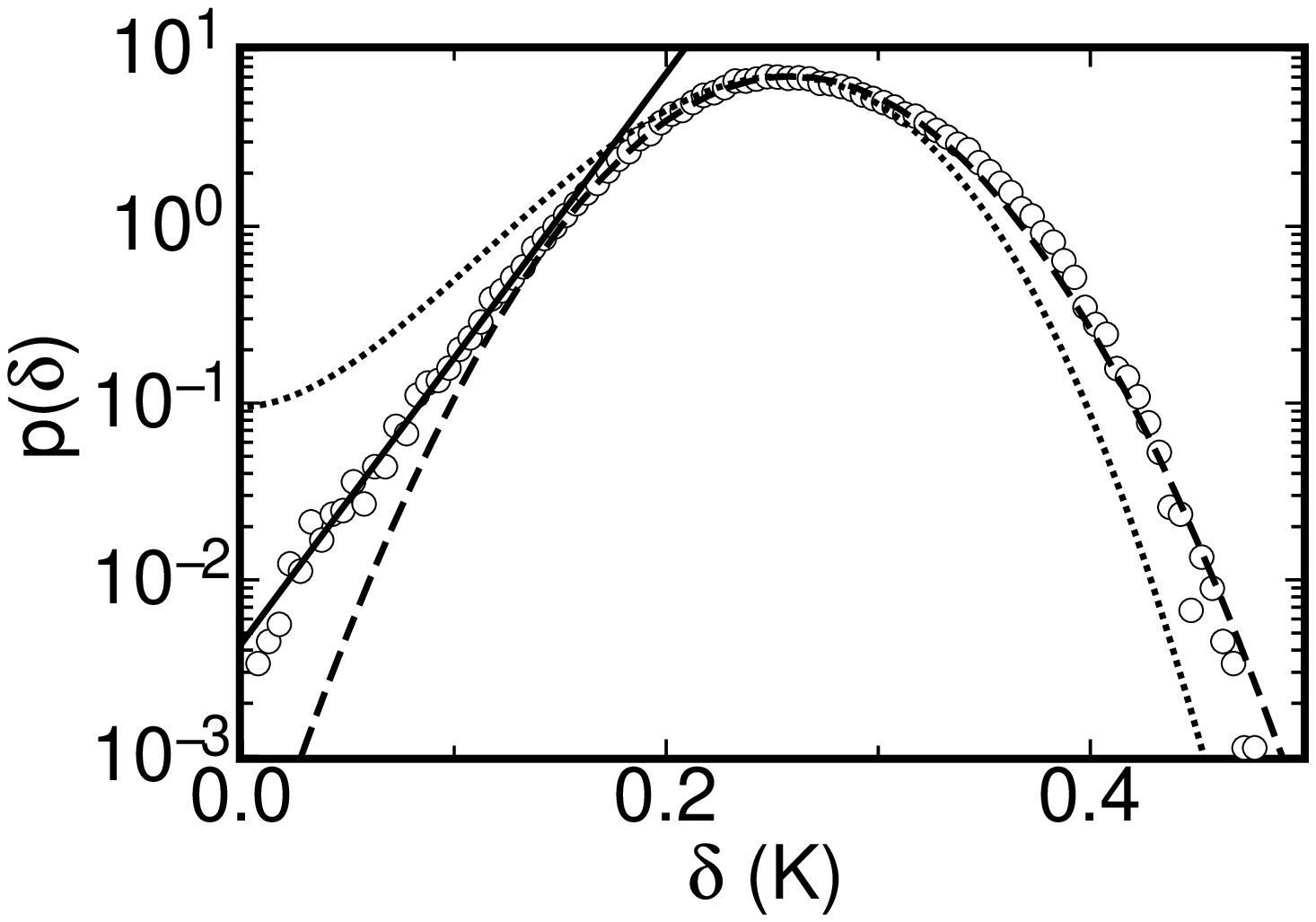}}
 \centerline{Fig. 5, Eric Brown, Physics of Fluids.}
\end{figure}

\begin{figure}                                                
\centerline{\includegraphics[width=3.25in]{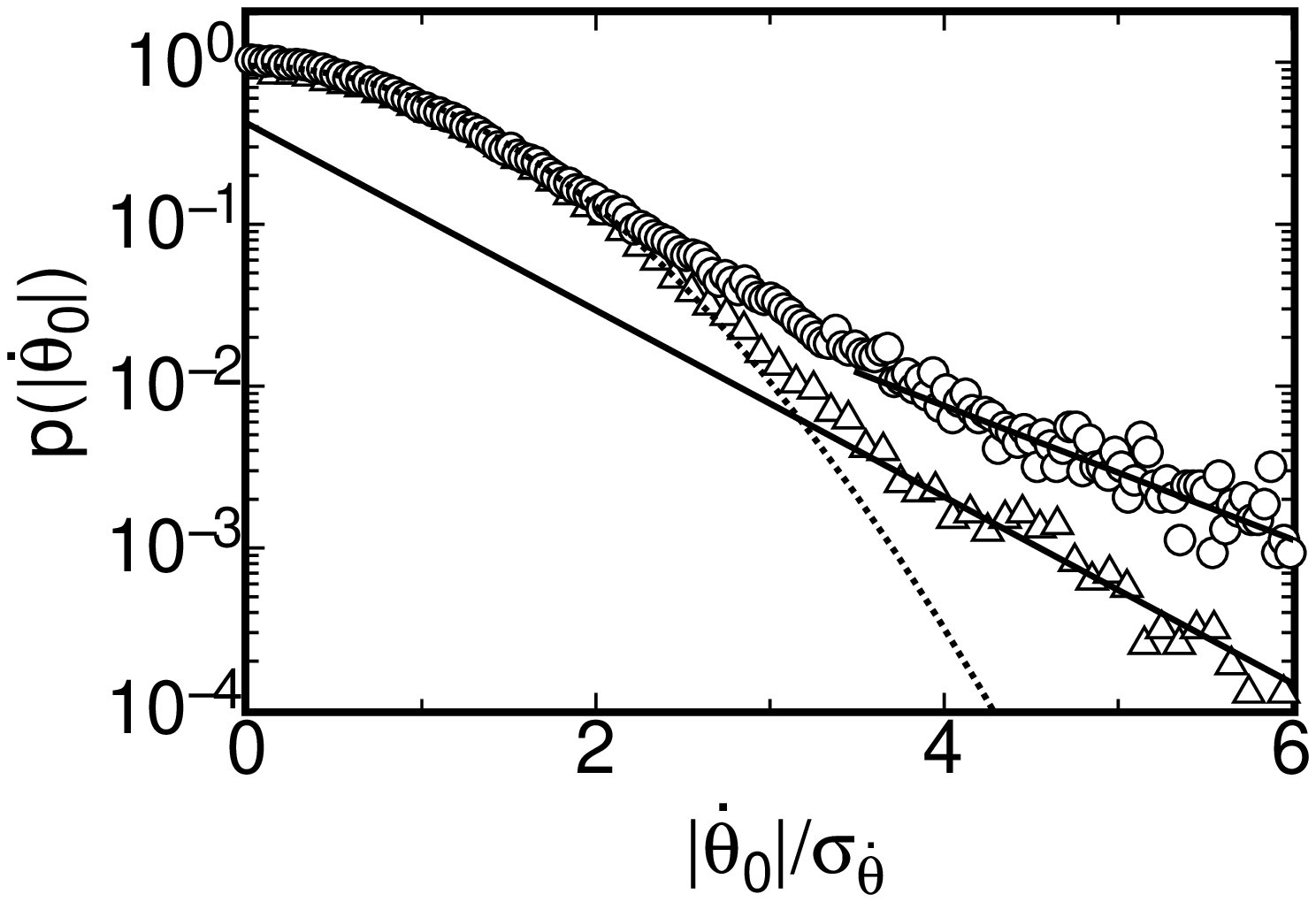}}
 \centerline{Fig. 6, Eric Brown, Physics of Fluids.}  
\end{figure}

\begin{figure}                                                
\centerline{\includegraphics[width=3.25in]{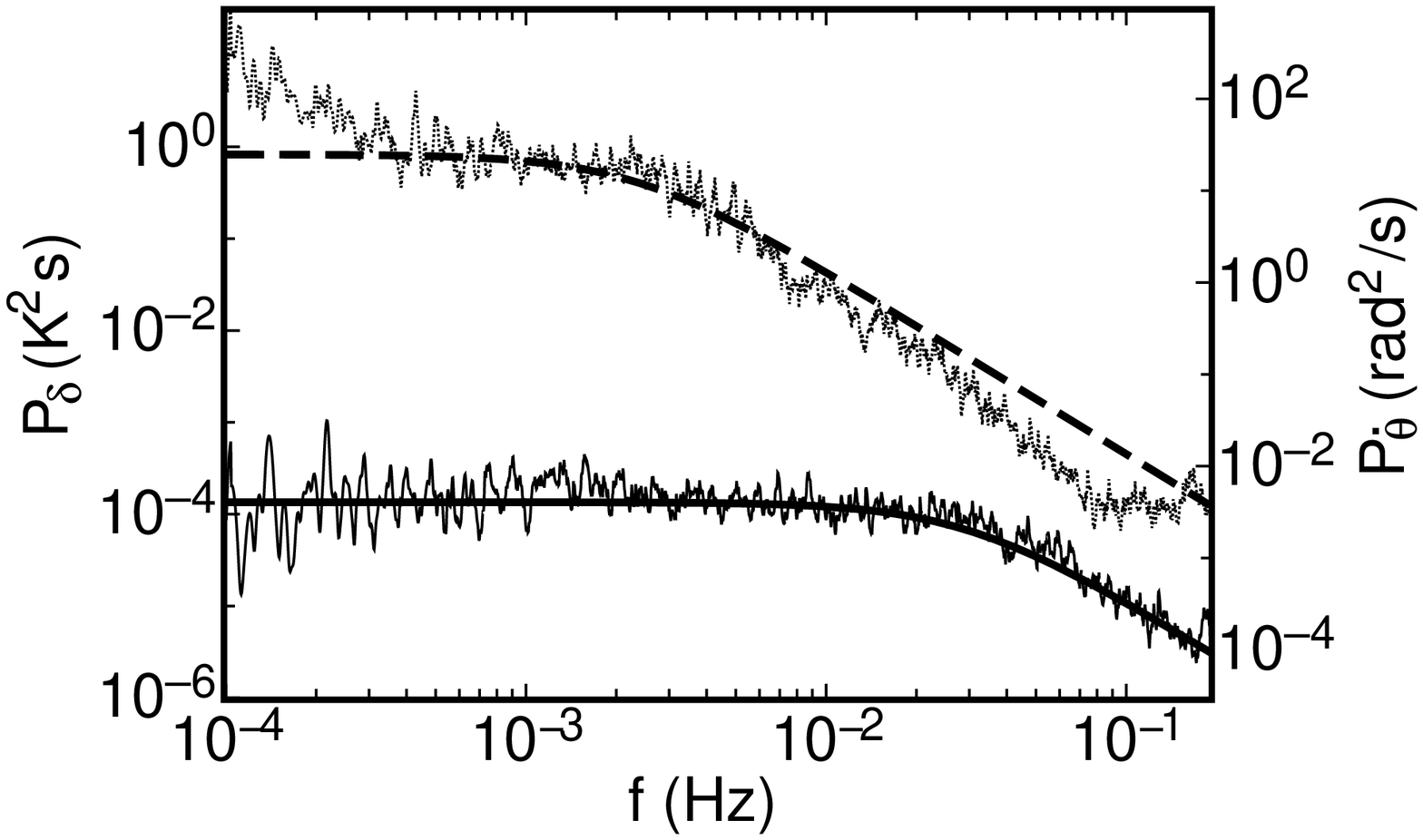}}
 \centerline{Fig. 7, Eric Brown, Physics of Fluids.}  
\end{figure}

\begin{figure}                                                
\centerline{\includegraphics[width=3.25in]{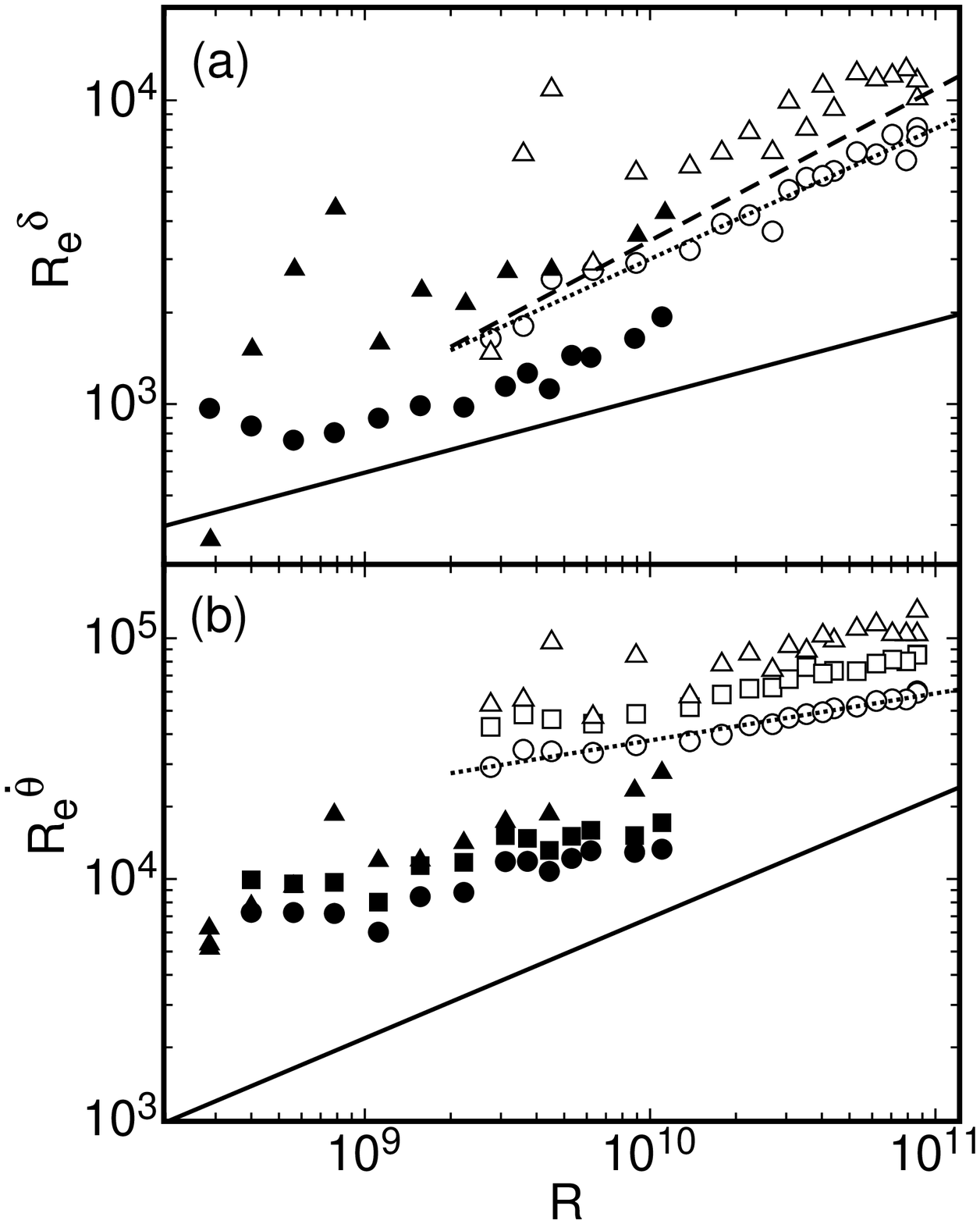}}
 \centerline{Fig. 8, Eric Brown, Physics of Fluids.}  
\end{figure}

 \begin{figure}
\center{ \includegraphics[width=3.25in]{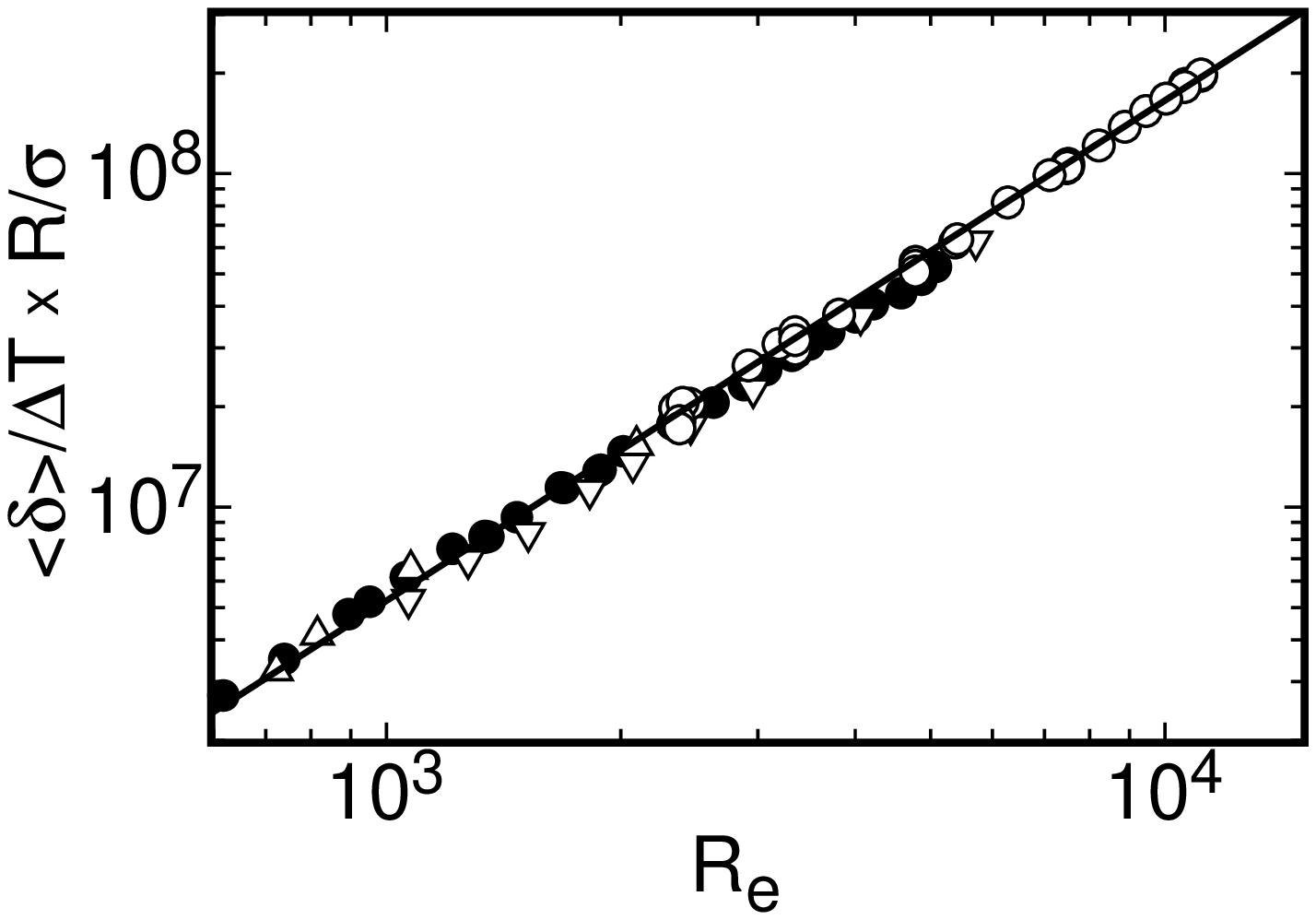}}
 \centerline{Fig. 9, Eric Brown, Physics of Fluids.}
\end{figure}

\begin{figure}                                                
\centerline{\includegraphics[width=3.25in]{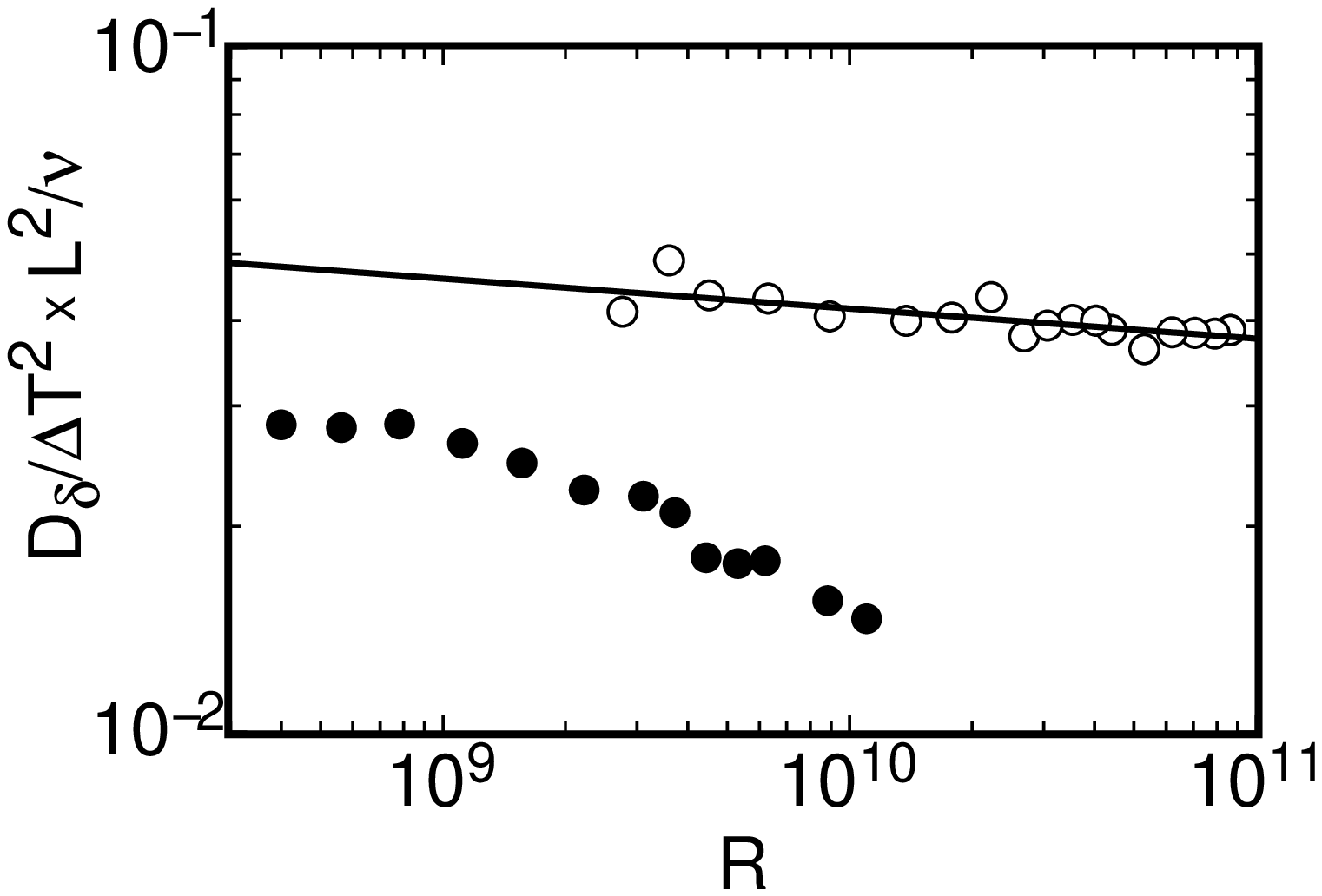}}
 \centerline{Fig. 10, Eric Brown, Physics of Fluids.}  
\end{figure}

\begin{figure}                                                
\centerline{\includegraphics[width=3.25in]{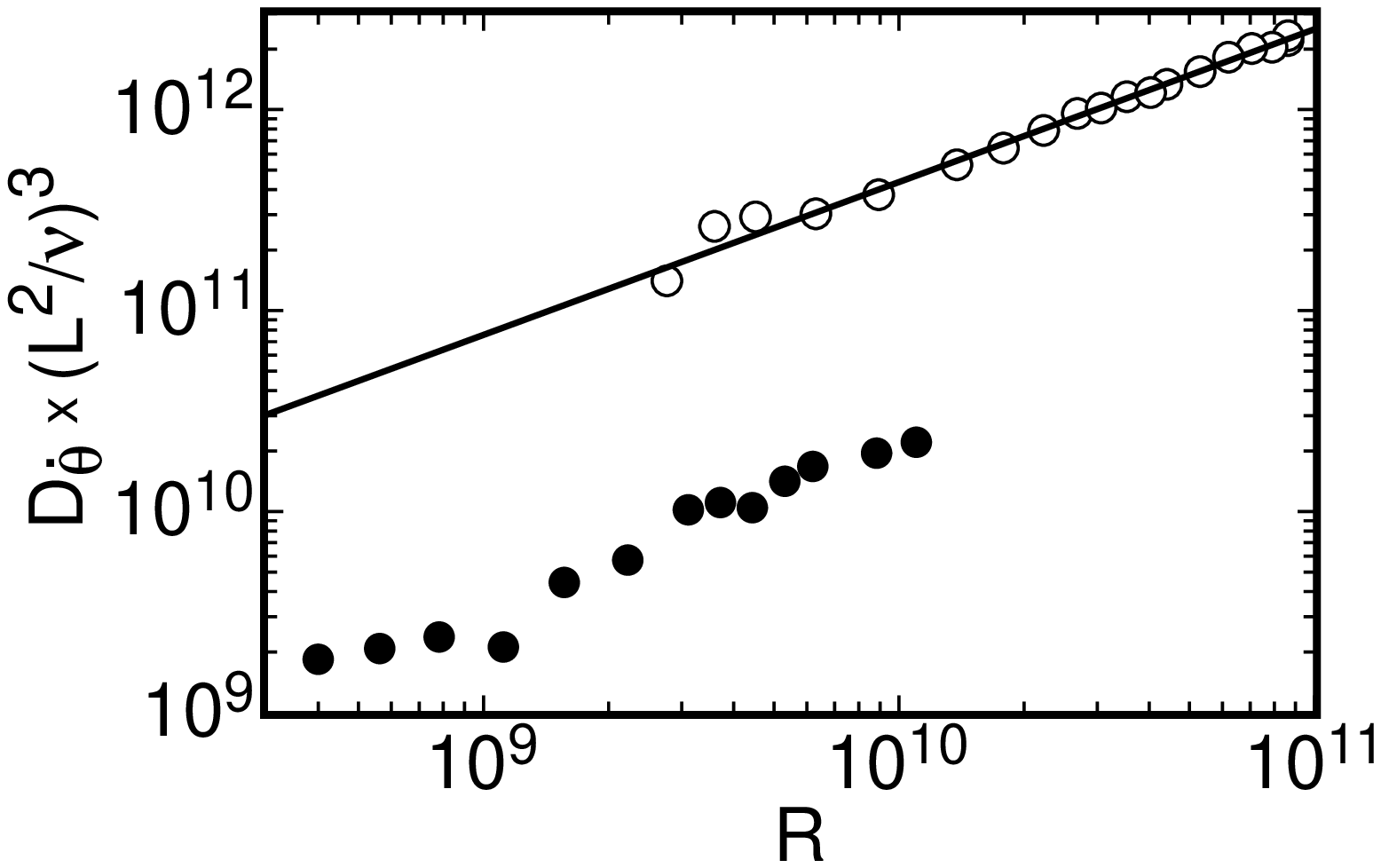}}
 \centerline{Fig. 11, Eric Brown, Physics of Fluids.}  
\end{figure}

\begin{figure}                                                
\centerline{\includegraphics[width=3.25in]{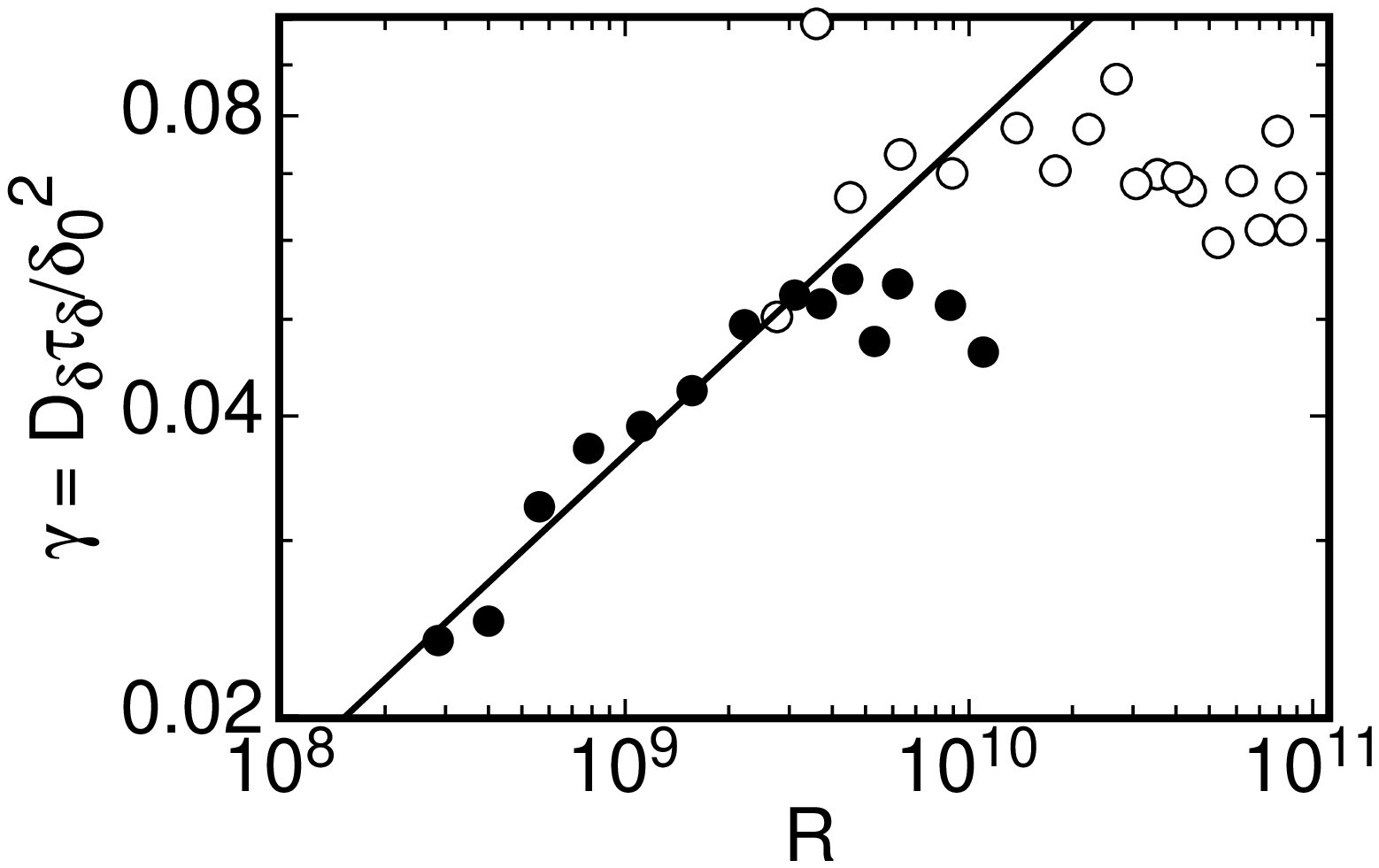}}
 \centerline{Fig. 12, Eric Brown, Physics of Fluids.}  
\end{figure}

\begin{figure}                                                
\centerline{\includegraphics[width=3.25in]{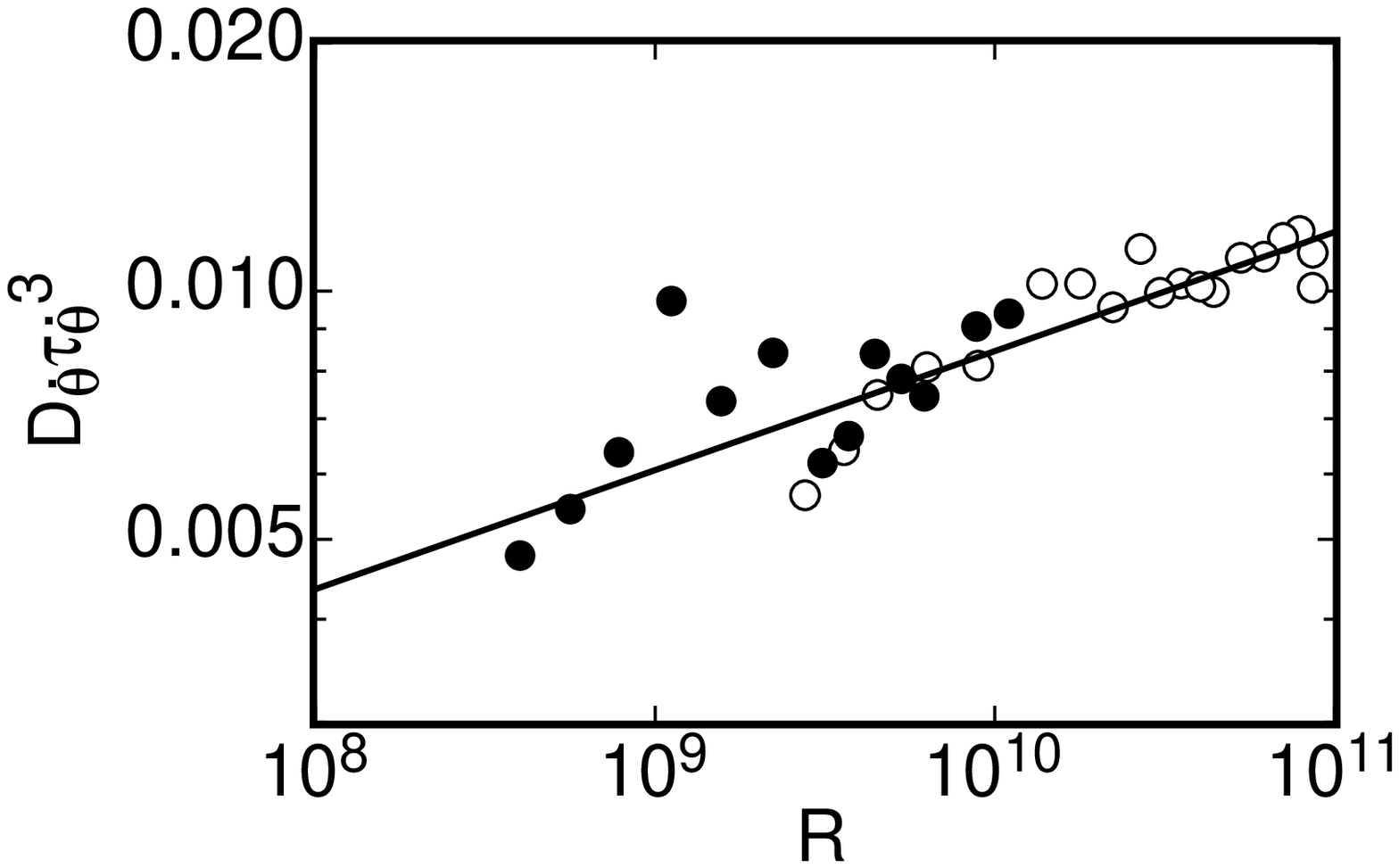}}
 \centerline{Fig. 13, Eric Brown, Physics of Fluids.}  
\end{figure}

 \begin{figure}
\center{ \includegraphics[width=3.25in]{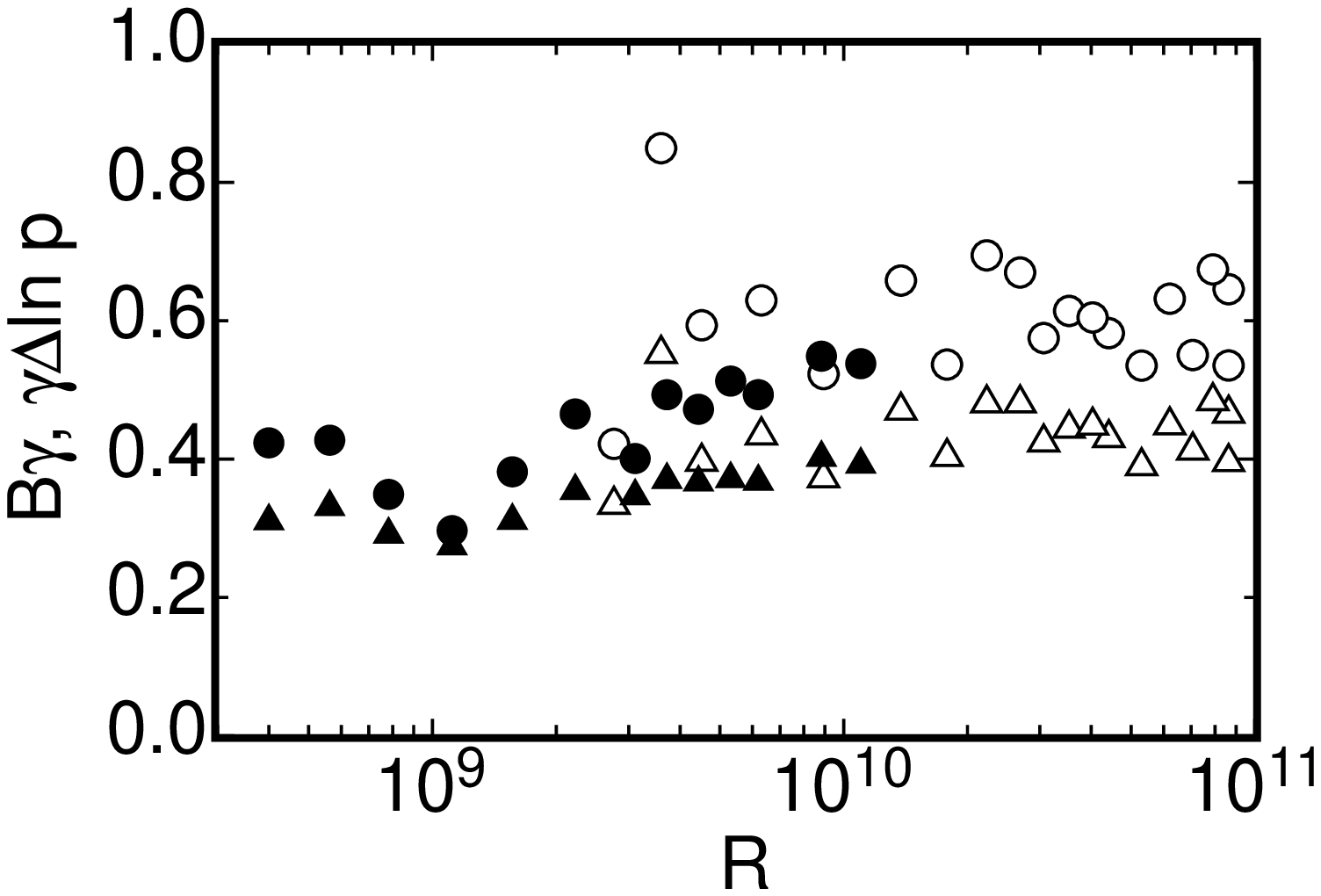}}
 \centerline{Fig. 14, Eric Brown, Physics of Fluids.}
\end{figure} 

\begin{figure}                                                
\centerline{\includegraphics[width=3.25in]{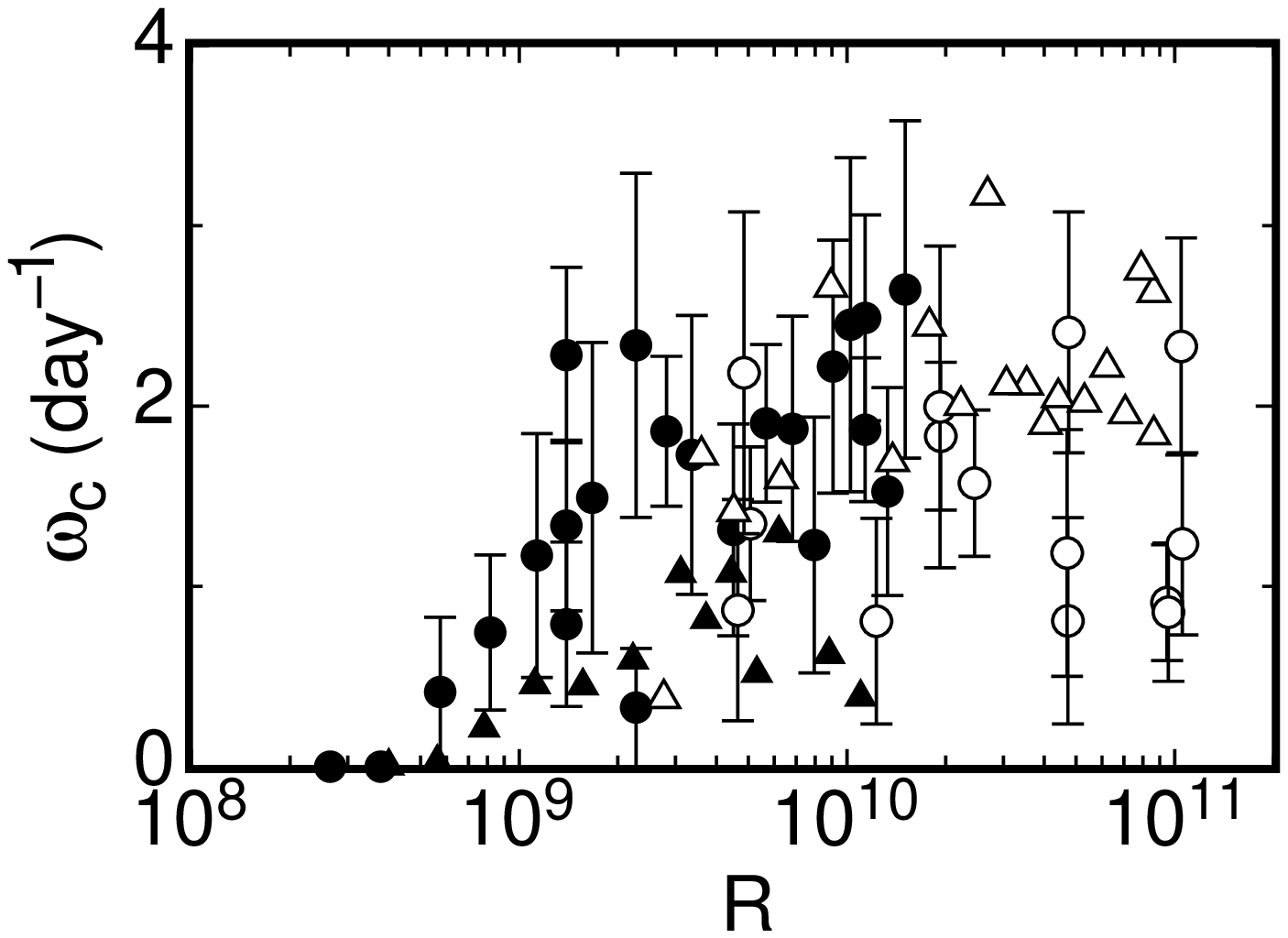}}
 \centerline{Fig. 15, Eric Brown, Physics of Fluids.}    
\end{figure}

\begin{figure}                                                
\centerline{\includegraphics[width=3.25in]{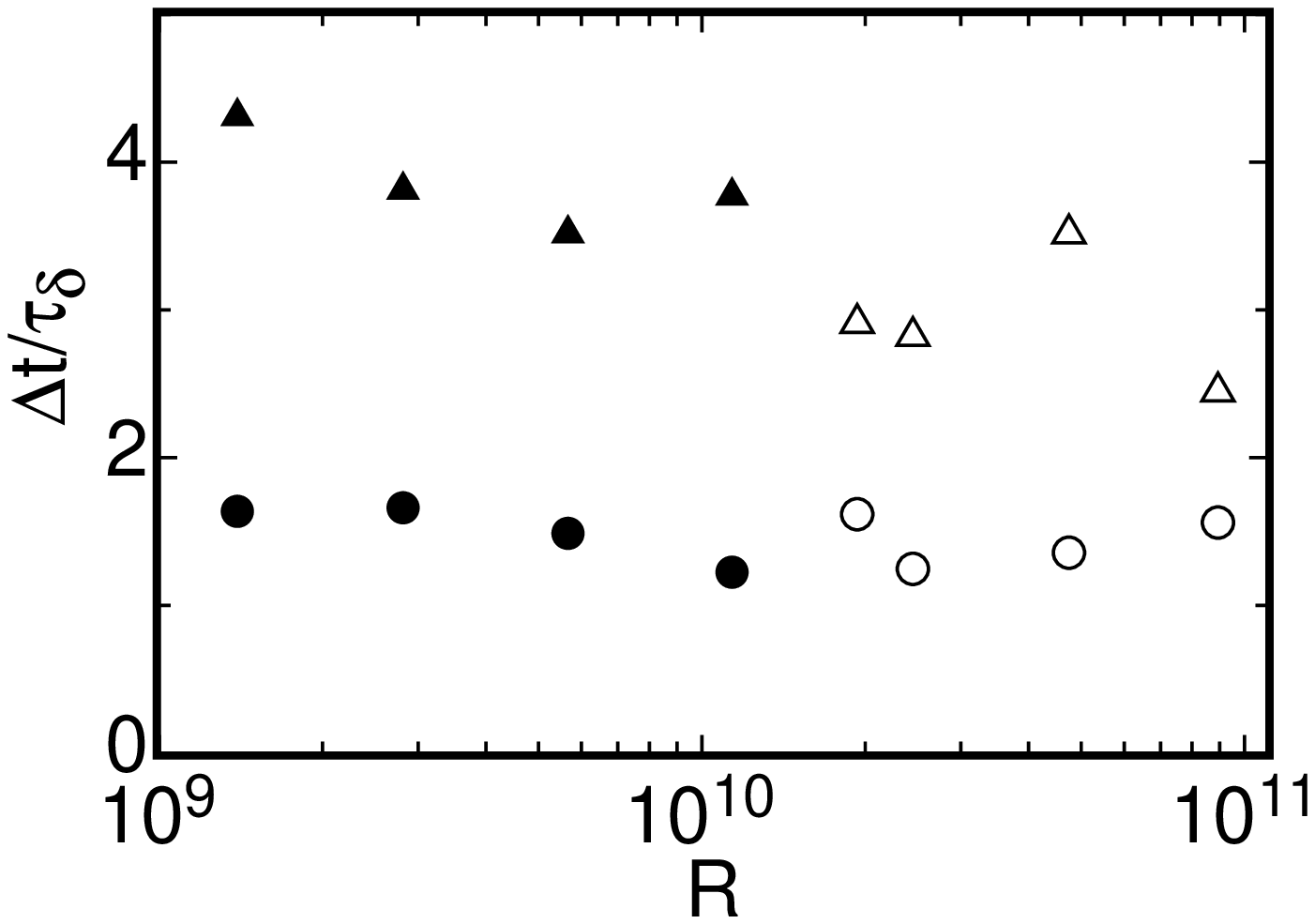}}
 \centerline{Fig. 16, Eric Brown, Physics of Fluids.}
\end{figure}

 \begin{figure}
\center{ \includegraphics[width=3.25in]{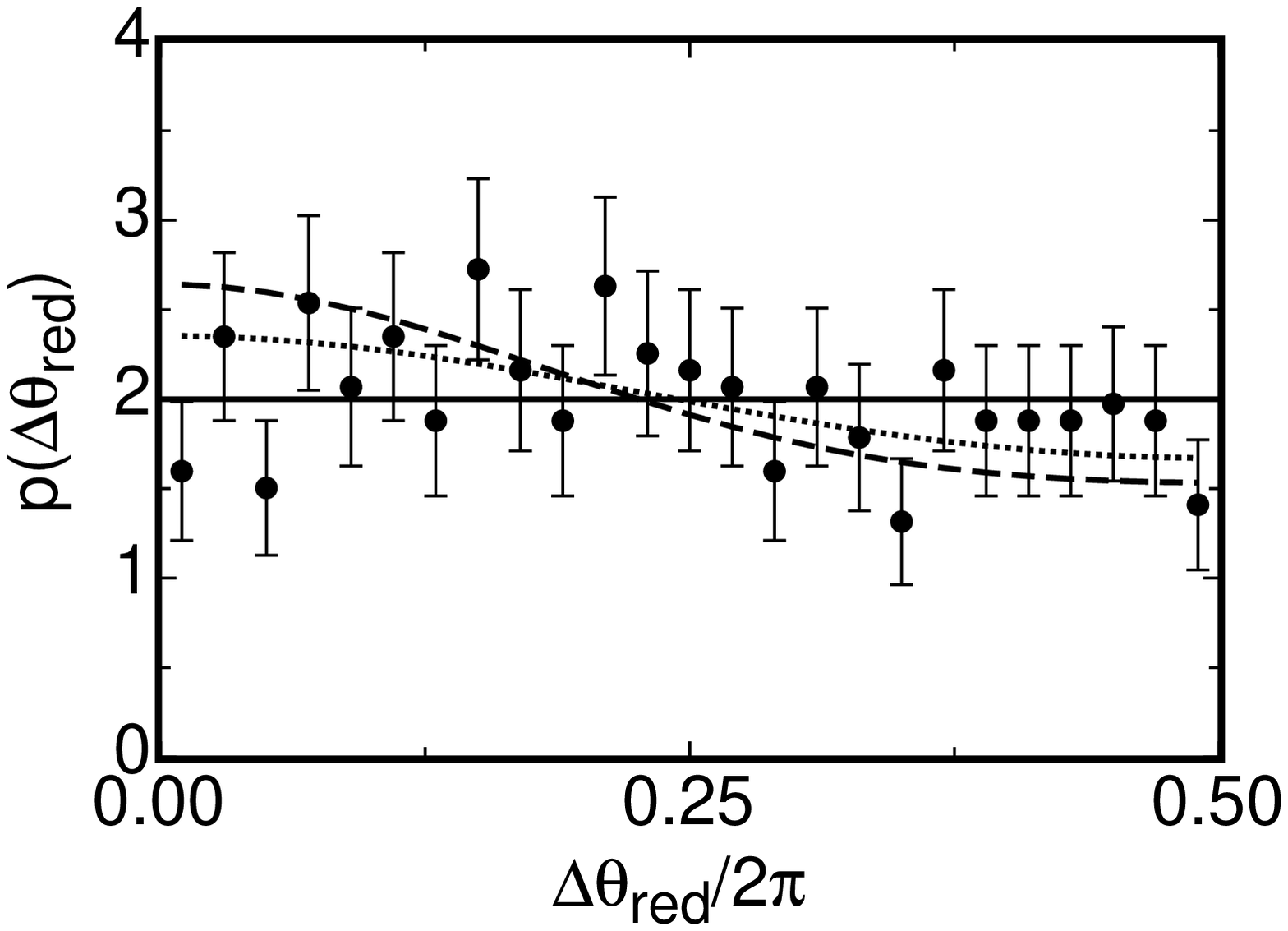}}
 \centerline{Fig. 17, Eric Brown, Physics of Fluids.}
\end{figure}


\begin{thebibliography}{10}

\bibitem{Si94}
E.~D. Siggia, {\em High {{Rayleigh}} number convection}, Annu. Rev. Fluid Mech.
  {\bf 26},  137  (1994).

\bibitem{Ka01}
L.~P. Kadanoff, {\em Turbulent heat flow: Structures and scaling}, Phys. Today
  {\bf 54},  34  (2001).

\bibitem{AGL02}
G. Ahlers, S. Grossmann, and D. Lohse, {\em Hochpr\"azision im {{Kochtopf}}:
  Neues zur turbulenten {{Konvektion}}}, Physik Journal {\bf 1},  31  (2002).

\bibitem{KH81}
R. Krishnamurti and L.~N. Howard, {\em Large scale flow generation in turbulent
  convection}, Proc. Natl. Acad. Sci. {\bf 78},  1981  (1981).

\bibitem{SWL89}
M. Sano, X.~Z. Wu, and A. Libchaber, {\em Turbulence in helium-gas free
  convection}, Phys. Rev. A {\bf 40},  6421  (1989).

\bibitem{CGHKLTWZZ89}
B. Castaing, G. Gunaratne, F. Heslot, L. Kadanoff, A. Libchaber, S. Thomae,
  X.~Z. Wu, S. Zaleski, and G. Zanetti, {\em Scaling of hard thermal turbulence
  in {{Rayleigh-B\'enard}} convection}, J. Fluid Mech. {\bf 204},  1  (1989).

\bibitem{CCL96}
S. Ciliberto, S. Cioni, and C. Laroche, {\em Large-scale flow properties of
  turbulent thermal convection}, Phys. Rev. E {\bf 54},  R5901  (1996).

\bibitem{QT01a}
X.~L. Qiu and P. Tong, {\em Large scale velocity structures in turbulent
  thermal convection}, Phys. Rev. E {\bf 64},  036304  (2001).

\bibitem{FA04}
D. Funfschilling and G. Ahlers, {\em Plume motion and large scale circulation
  in a cylindrical {{Rayleigh-B\'enard}} cell}, Phys. Rev. Lett. {\bf 92},
  194502  (2004).

\bibitem{SXT05}
C. Sun, K.~Q. Xia, and P. Tong, {\em Three-dimensional flow structures and
  dynamics of turbulent thermal convection in a cylindrical cell}, Phys. Rev. E
  {\bf 72},  026302  (2005).

\bibitem{TMMS05}
Y. Tsuji, T. Mizuno, T. Mashiko, and M. Sano, {\em Mean Wind in Convective
  Turbulence of Mercury}, Phys. Rev. Lett. {\bf 94},  034501  (2005).

\bibitem{HCL87}
F. Heslot, B. Castaing, and A. Libchaber, {\em Transition to turbulence in
  helium gas}, Phys. Rev. A {\bf 36},  5870  (1987).

\bibitem{TSGS96}
T. Takeshita, T. Segawa, J.~A. Glazier, and M. Sano, {\em Thermal turbulence in
  mercury}, Phys. Rev. Lett. {\bf 76},  1465  (1996).

\bibitem{CCS97}
S. Cioni, S. Ciliberto, and J. Sommeria, {\em Strongly turbulent
  {{Rayleigh-B\'enard}} convection in mercury: comparison with results at
  moderate {{Prandtl}} number}, J. Fluid Mech. {\bf 335},  111  (1997).

\bibitem{QYT00}
X.~L. Qiu, S.~H. Yao, and P. Tong, {\em Large-scale coherent rotation and
  oscillation in turbulent thermal convection}, Phys. Rev. E {\bf 61},  R6075
  (2000).

\bibitem{QT01b}
X.~L. Qiu and P. Tong, {\em Onset of coherent oscillations in turbulent
  {{Rayleigh-B\'enard}} convection}, Phys. Rev. Lett {\bf 87},  094501  (2001).

\bibitem{NSSD01}
J. Niemela, L. Skrbek, K.~R. Sreenivasan, and R.~J. Donnelly, {\em The wind in
  confined thermal turbulence}, J. Fluid Mech. {\bf 449},  169  (2001).

\bibitem{QT02}
X.~L. Qiu and P. Tong, {\em Temperature oscillations in turbulent
  {{Rayleigh-B\'enard}} convection}, Phys. Rev. E {\bf 66},  026308  (2002).

\bibitem{QSTX04}
X.~L. Qiu, X.~D. Shang, P. Tong, and K.-Q. Xia, {\em Velocity oscillations in
  turbulent {{Rayleigh-B\'enard}} convection}, Phys. Fluids. {\bf 16},  412
  (2004).

\bibitem{FBA08}
D. Funfschilling, E. Brown, and G. Ahlers, {\em Azimuthal oscillations of the
  large-scale circulation in turbulent {{Rayleigh-B\'enard}} convection}, J.
  Fluid Mech  (2008), in print.

\bibitem{SXX05}
C. Sun, H.~D. Xi, and K.~Q. Xia, {\em Azimuthal symmetry, flow dynamics, and
  heat transport in turbulent thermal convection in a cylinder with an aspect
  ratio of 0.5}, Phys. Rev. Lett. {\bf 95},  074502  (2005).

\bibitem{XZX06}
H.~D. Xi, Q. Zhou, and K.~Q. Xia, {\em Azimuthal motion of the mean wind in
  turbulent thermal convestion}, Phys. Rev. E {\bf 73},  056312  (2006).

\bibitem{BA06a}
E. Brown and G. Ahlers, {\em Rotations and cessations of the large-scale
  circulation in turbulent {{Rayleigh-B{\'e}nard}} convection}, J. Fluid Mech.
  {\bf 568},  351  (2006).

\bibitem{BA06b}
E. Brown and G. Ahlers, {\em Effect of the {{Earth's}} {{Coriolis}} force on
  turbulent {{Rayleigh-B{\'e}nard}} convection in the laboratory}, Phys. Fluids
  {\bf 18},  125108  (2006).

\bibitem{BNA05}
E. Brown, A. Nikolaenko, and G. Ahlers, {\em Reorientation of the large-scale
  circulation in turbulent {{Rayleigh-B{\'e}nard}} convection}, Phys. Rev. Lett
  {\bf 95},  084503  (2005).

\bibitem{GCHR99}
G. Glatzmaier, L.~H. R.~Coe, and P. Roberts, {\em The role of the {{Earth}}'s
  mantle in controlling the frequency of geomagnetic reversals}, Nature(London)
  {\bf 401},  885  (1999).

\bibitem{DDSC00}
E. van Doorn, B. Dhruva, K.~R. Sreenivasan, and V. Cassella, {\em Statistics of
  wind direction and its increments}, Phys. Fluids {\bf 12},  1529  (2000).

\bibitem{HL80b}
R. Howard and B. LaBonte, {\em The Sun is observed to be a torsional oscillator
  with a period of 11 years}, Astrophys. J. {\bf 239},  L33  (1980).
).

\bibitem{SBN02}
K.~R. Sreenivasan, A. Bershadski, and J. Niemela, {\em Mean wind and its
  reversals in thermal convection}, Phys. Rev. E {\bf 65},  056306  (2002).

\bibitem{Be05}
R. Benzi, {\em Flow reversal in a simple dynamical model of turbulence}, Phys.
  Rev. Lett. {\bf 95},  024502  (2005).

\bibitem{FGL05}
F. {{Fontenele Araujo}}, S. Grossmann, and D. Lohse, {\em Wind reversals in
  turbulent {{Rayleigh-B\'enard}} convection}, Phys. Rev. Lett. {\bf 95},
  084502  (2005).

\bibitem{RPTDGFL06}
Resagk, R. du~Puits, , A. Thess, F. Dolzhansky, S. Grossmann, F. {Fontenele
  Araujo}, and D. Lohse, {\em Oscillations of the large scale wind in turbulent
  thermal convection}, Phys. Fluids {\bf 18},  095105  (2006).


\bibitem{BA07a}
E. Brown and G. Ahlers, {\em Large-scale circulation model of turbulent
  {{Rayleigh-B{\'e}nard}} convection}, Phys. Rev. Lett. {\bf 98},  134501
  (2007).

\bibitem{BNFA05}
E. Brown, A. Nikolaenko, D. Funfschilling, and G. Ahlers, {\em Heat transport
  in turbulent {{Rayleigh-B\'enard}} convection: Effect of finite top- and
  bottom-plate conductivities}, Phys. Fluids {\bf 17},  075108  (2005).

\bibitem{BA07b}
E. Brown and G. Ahlers, {\em Temperature gradients, and search for
  non-{{Boussinesq}} effects, in the interior of turbulent {{Rayleigh-B\'enard}}  convection}, 
  Europhys. Lett. {\bf 80},  14001  (2007).

\bibitem{ABN06}
G. Ahlers, E. Brown, and A. Nikolaenko, {\em The search for slow transients,
  and the effect of imperfect vertical alignment, in turbulent
  {{Rayleigh-B\'enard}} convection}, J. Fluid Mech. {\bf 557},  347  (2006).



\bibitem{GL02}
S. Grossmann and D. Lohse, {\em Prandtl and {{Rayleigh}} number dependence of
  the {{Reynolds}} number in turbulent thermal convection}, Phys. Rev. E {\bf
  66},  016305  (2002).


\bibitem{ABFFGL06}
G. Ahlers, E. Brown, F. {{Fontenele Araujo}}, D. Funfschilling, S. Grossmann,
  and D. Lohse, {\em Non-{{Oberbeck-Boussinesq}} effects in strongly turbulent
  {{Rayleigh-B\'enard}} convection}, J. Fluid Mech. {\bf 569},  409  (2006).

\bibitem{AFFGL07}
G. Ahlers, F. {{Fontenele Araujo}}, D. Funfschilling, S. Grossmann,
  and D. Lohse, {\em Non-{{Oberbeck-Boussinesq}} effects in gaseous
  {{Rayleigh-B\'enard}} convection}, Phys. Rev. Lett. {\bf 98}, 054501 (2007).

\bibitem{Tong_private}
P. Tong, private communication, 2007.

\bibitem{Kaz}
K. Sugiyama, private communication, 2007.

\bibitem{Kr40}
H.~A. Kramers, {\em Brownian motion in a field of force and the diffusion model
  of chemical reactions}, Physica (Amsterdam) {\bf 7},  284  (1940).

\bibitem{BFA07}
E. Brown, D. Funfschilling, and G. Ahlers, {\em Anomalous {R}eynolds-number
  scaling in turbulent {{Rayleigh-B{\'e}nard}} convection}, J. Stat. Mech.
  {P}10005  (2007).

\bibitem{Gi05}
M. Gitterman, {\em The Noisy Oscillator, The First Hundred Years, From Einstein
  Until Now} (World Scientific, Singapore, 2005).

\bibitem{XX07}
H.~D. Xi and K.~Q. Xia, {\em Cessations and reversals of the large-scale
  circulation in turbulent thermal convection}, Phys. Rev. E {\bf 75},  066307
  (2007).

\bibitem{NBFA05}
A. Nikolaenko, E. Brown, D. Funfschilling, and G. Ahlers, {\em Heat transport
  by turbulent {{Rayleigh-B\'enard}} convection in cylindrical cells with
  aspect ratio one and less}, J. Fluid Mech. {\bf 523},  251  (2005).

\end{thebibliography}
\end{document}